\documentclass[twocolumn,noshowpacs,superscriptaddress]{revtex4-2}
\usepackage{bm}
\usepackage{siunitx}
\usepackage{amsmath, amsfonts, amssymb}
\usepackage{diagbox}
\usepackage[usenames]{color}
\usepackage{colortbl}
\usepackage{booktabs}
\usepackage{multirow}
\usepackage[colorlinks=true, linkcolor=black, urlcolor=blue, citecolor=blue]{hyperref}
\usepackage{color,soul}
\usepackage{ulem}
\usepackage{graphicx, epstopdf, subfigure}
\usepackage{float}
\usepackage{xr}
\usepackage{braket}

\usepackage{color, colortbl}
\definecolor{Gray}{gray}{0.9}
\definecolor{LightCyan}{rgb}{0.88,1,1}

\begin{document}

\title{Study of atomic effects on electron spectrum in bound-muon decay process}

\author{M.\,Y.~Kaygorodov}
\affiliation{School of Physics and Information Technology, Shaanxi Normal University, Xi'an 710119, China}
\affiliation{Department of Physics, University of Alberta, Edmonton, Alberta, Canada T6G 2G7}

\author{Y.\,S.~Kozhedub}
\affiliation{Department of Physics, St.~Petersburg State University, Universitetskaya 7-9, 199034 St.~Petersburg, Russia\looseness=-1}

\author{A.\,V.~Malyshev}
\affiliation{Department of Physics, St.~Petersburg State University, Universitetskaya 7-9, 199034 St.~Petersburg, Russia\looseness=-1}
\affiliation{Petersburg Nuclear Physics Institute named by B.P. Konstantinov of National Research Center ``Kurchatov Institute'', Orlova roscha 1, 188300 Gatchina, Leningrad region, Russia}

\author{A.\,O.~Davydov}
\affiliation{Department of Physics, University of Alberta, Edmonton, Alberta, Canada T6G 2G7}

\author{Y. Wu}
\affiliation{Institute of Applied Physics and Computational Mathematics, Beijing 100088, China}

\author{S. B. Zhang}
\affiliation{School of Physics and Information Technology, Shaanxi Normal University, Xi'an 710119, China}

\begin{abstract}
For the bound-muon decay process, the study of atomic effects on the electron spectrum near its endpoint is performed within the framework of the Fermi effective theory.
The analysis takes into account for corrections due to finite-nuclear-size, nuclear-deformation, electron-screening, and vacuum-polarization effects, all of which are incorporated self-consistently into the Dirac equation.
Furthermore, the nuclear-recoil correction to the muon binding energy is included.
Calculations are carried out for the isotopes of C, Al, and Si, which are of a particular importance for forthcoming experiments aimed at search for the charged-lepton flavor-violating process of muon-to-electron conversion in a nuclear field.
\end{abstract}

\maketitle

\section{Introduction}
In contrast to the well-established phenomena of quark-generation mixing and neutrino-flavor oscillations, there is still no experimental evidence for the charged-lepton flavor violation~(CLFV), see, e.g., the reviews in Refs.~\cite{2008_MarcianoW_AnnRevNuclPartPhys58, 2013_BernsteinR_PR532, 2018_CalibbiL_NouvCim41, 2022_ArduM_MDPI8} and references therein.
One of promising channels for investigating the CLFV is the coherent muon-to-electron conversion in the field of a nucleus.
As this process may involve hypothetical lepton-quark interactions, it can be used to test various extensions of the Standard Model.
The current experimental upper limit on the conversion branching ratio is reported in Ref.~\cite{2006_BertlW_EPJC47}.
Several next-generation experiments~\cite{2019_TeshimaN_arXiv1911, 2022_MoritsuM_MDPI8, 2025_MiscettiS_NIMPA1073}, based on the grounds of Ref.~\cite{1989_DzhilkibaevR_YadFiz49}, aim to significantly improve the sensitivity to this rare process.
The prospects of these experiments to explore physics beyond the Standard Model are discussed in Ref.~\cite{2024_HillR_PRD109}.
\par
The bound-muon decay, also referred to in the literature as the~$\mu^-$ decay in orbit , is a source of physical background which accompanies the muon-to-electron conversion.
The experiments on the conversion are focused on the electron energy spectrum near its endpoint, where the background contributions are expected to be minimal.
However, in the case of bound-muon decay, the electron spectrum close to the endpoint is strongly affected by atomic effects.
Consequently, to reliably identify the muon-to-electron conversion signal, it is essential to develop a precise theoretical model of the Standard Model-allowed background arising from bound-muon decay, with careful inclusion of all relevant atomic effects.
\par
The theoretical study of the bound-muon decay process was initiated in Refs.~\cite{1949_TiomnoJ_RMP21, 1951_PorterC_PR83}.
Later, various aspects of the problem were investigated in Refs.~\cite{1959_TenagliaL_NouvCim13, 1960_GilinskyV_PR120, 1960_UeberallH_PR119, 1961_HuffR_AnnPhys16, 1961_JohnsonW_PR124, 1979_vonBaeyerH_PRA19, 1974_HaenggiP_PhysLettB51, 1980_HerzogF_HevlPhysActa53,  1980_ChatterjeeL_AnnDerPhys492, 1982_ShankerO_PRD25}.
The most recent and comprehensive analysis of the complete electron spectrum across different nuclei was carried out in Refs.~\cite{1987_WatanabeR_ProgTheorPhys78, 1993_WatanabeR_ADNDT54}, where fully relativistic wave functions constructed for the finite-nuclear-charge distribution model were employed.
The specific case of~${}^{27}_{13}\mathrm{Al}$ was scrutinized in Refs.~\cite{2011_CzarneckiA_PRD84, 2016_SzafronR_PhysLettB753, 2016_SzafronR_PRD94}.
The isotope dependence of the electron spectrum near the endpoint was explored in Ref.~\cite{2022_HeeckJ_PRD105}.
\par
The present study has two primary goals.
The first objective is to derive a general expression for the electron spectrum corresponding to an arbitrary bound state of the muon.
The second one is to study the influence of various atomic effects, including finite nuclear size, nuclear deformation, muon nuclear recoil, and electron screening, on both the muon binding energy and the resulting electron spectrum.
In addition, corrections arising from quantum electrodynamics (QED), specifically the vacuum-polarization effect, are also considered.
The analysis of atomic effects on the electron spectrum is performed for several isotopes:~${}^{12}_{\,\,\,6}\mathrm{C}$ and~${}^{28}_{14}\mathrm{Si}$, which will be used in the experiment described in Ref.~\cite{2019_TeshimaN_arXiv1911}, and~${}^{27}_{13}\mathrm{Al}$, which will be employed in the experiments discussed in Refs.~\cite{2022_MoritsuM_MDPI8, 2025_MiscettiS_NIMPA1073}.
\par
The paper is organized as follows.
In Sec.~\ref{sec:2_theory}, a brief derivation of a general expression for the electron spectrum within the central-field approximation is presented, employing two independent but equivalent approaches to verify the consistency of the results.
Sec.~\ref{sec:3_results} outlines the details of the numerical calculations, presents and discusses the results, and provides a comparison with existing data from the literature.
Sec.~\ref{sec:4_conclusion} is reserved for the conclusions.
The manuscript also contains seven appendices, each addressing specific technical aspects related to the derivations in Sec.~\ref{sec:2_theory}.
\par
The relativistic units are used throughout the paper unless specified otherwise.

\section{Theory}\label{sec:2_theory}
The bound-muon decay process is described by the reaction
\begin{equation}\label{eq:muon_decay_reaction}
    \mu^{-}\to e^{-}+\bar{\nu}_{e}+\nu_{\mu},
\end{equation}
where the initial-state muon~$\mu^{-}$ is bound in a spherically symmetric nuclear potential, final-state electron~$e^-$ is unbound, and the final-state neutrinos~$\bar{\nu}_{e}$ and~$\nu_{\mu}$ can be regarded as free massless particles.
Instead of treating the reaction as a conventional three-body decay, it is advantageous to reformulate it as an equivalent two-to-two scattering process:
\begin{equation}\label{eq:muon_decay_scattering}
    \mu^{-}+\nu_{e}\left(p_{\nu_e}\right)\to e^{-}+\nu_{\mu}\left(p_{\nu_\mu}\right),
\end{equation}
where the outgoing electron anti-neutrino~$\bar{\nu}_e$, characterized by an asymptotic four-momentum~$p_{\bar{\nu}_e}$, is reinterpreted as an incoming neutrino with opposite four-momentum~$p_{\nu_e}=-p_{\bar{\nu}_e}$.
In Eq.~\eqref{eq:muon_decay_scattering},~$p_{\nu_{\mu}}$ is an asymptotic four-momentum of the muonic neutrino~$\nu_{\mu}$, and its definition stays unchanged compared to Eq.~\eqref{eq:muon_decay_reaction}.
This formulation facilitates the application of standard in quantum-field-theory techniques.
\par
Within the framework of the Fermi effective theory, the lepton-neutrino interaction operator governing the process described in Eq.~\eqref{eq:muon_decay_scattering} takes the following form:
\begin{equation}\label{eq:fermi_interaction_operator}
\begin{aligned}V^{\mathrm{F}}\left(1,2\right)= & \frac{G_{\mathrm{F}}}{\sqrt{2}}\delta\left(\vec{r}_{1}-\vec{r}_{2}\right)\\
\times & \left[\gamma^{0}\gamma^{\rho}\left(1-\gamma^{5}\right)\right]\left(1\right)\left[\gamma^{0}\gamma_{\rho}\left(1-\gamma^{5}\right)\right]\left(2\right),
\end{aligned}
\end{equation}
where the indices~$(1)$ and~$(2)$ here and in what follows label the particles the operator acts on,~$G_{\mathrm{F}}$ is the Fermi constant, $\gamma^{\rho}$ are the Dirac gamma matrices, and the summation over the repeated Lorentz indices is implied.
The tree-level amplitude of the process is given by
\begin{equation}\label{eq:A}
    A=\braket{e\nu_\mu|V^\mathrm{F}|\mu\nu_e},
\end{equation}
where the bra-ket notation is employed and the written matrix element implies integration over the spatial and spin coordinates of the particles involved.
\par
Given the spherical symmetry of the system, it is natural to adopt the central-field approximation.
To obtain the electron spectrum, one has to integrate over the quantum numbers of neutrinos, which are not observed in the experiment, accounting for all their possible configurations consistent with the energy-conservation law,~$E_\mu+E_{\nu_e}=E_e+E_{\nu_\mu}$.
The operator in Eq.~\eqref{eq:fermi_interaction_operator} ensures that only neutrino states with the appropriate helicity contribute to the transition amplitude.
As a result, the integration domain over the neutrino variables in Eq.~\eqref{eq:A} can be formally extended without altering the outcome.
By employing different representations of the neutrino states in Eq.~\eqref{eq:A}, two different but formally equivalent approaches for evaluating the electron spectrum emerge.
These alternative formulations lead to different final expressions, the equivalence of which can be demonstrated explicitly.
For completeness, both approaches are presented and discussed in detail in this work.
\par
The first method is a general, brute-force approach in which all particles are represented in the spherical-wave basis.
For each particle, the basis functions are described by a set of relativistic central-field quantum numbers:~$\zeta\equiv \left(E, \varkappa,\mu \right)$, where~$E$ denotes the energy (or an equivalent quantum number),~$\varkappa$ is the relativistic angular quantum number, and~$\mu$ is the projection of the total angular momentum~$j=|\varkappa|-1/2$.
For the bound muon, the principal quantum number~$n$ is used instead of the energy~$E$.
The differential decay rate in this approach is given by
\begin{equation}
\begin{aligned}\frac{dW^{2\mathrm{p}}\left(\zeta_{e},\zeta_{\mu}\right)}{dE_{e}} & =2\pi\int\,\delta\left(-E_{e}-E_{\nu_{\mu}}+E_{\mu}+E_{\nu_{e}}\right)\\
 & \times\left|A\left(\zeta_{e},\zeta_{\nu_{\mu}},\zeta_{\mu},\zeta_{\nu_{e}}\right)\right|^{2} d\zeta_{\nu_{\mu}}\, d\zeta_{\nu_{e}},
\end{aligned}
\end{equation}
where it is assumed that all unbound sates are normalized to the delta function in energy.
This formulation will be referred below to as the \textit{two-particle approach} and labeled with ''2p``.
\par
The second approach represents only the muon and electron states in the spherical-wave basis, while the neutrinos are described using the plane-wave basis, characterized by a set of quantum numbers~$\xi=\left(\vec{p},m_s\right)$, where $\vec{p}$ is the three-momentum and~$m_s$ is the spin projection onto~$\vec{p}$.
The latter wave functions are normalized to the delta function in momentum space.
In this approach, the partial-wave electron spectrum can be expressed as
\begin{equation}
\begin{aligned}\frac{dW^{1\mathrm{p}}\left(\zeta_{e},\zeta_{\mu}\right)}{dE_{e}} & =2\pi\int\,\delta\left(-E_{e}-E_{\nu_{\mu}}+E_{\mu}+E_{\nu_{e}}\right)\\
 & \times\left|A\left(\zeta_{e},\xi_{\nu_{\mu}},\zeta_{\mu},\xi_{\nu_{e}}\right)\right|^{2} d\xi_{\nu_{\mu}}\, d\xi_{\nu_{e}},
\end{aligned}
\end{equation}
This mixed representation effectively reduces the two-particle problem to a one-particle problem.
Such a reduction is justified by the fact that the electromagnetic field of the nucleus, responsible for binding the muon, does not interact with the neutrinos.
Accordingly, this method will be referred to as the \textit{effective one-particle approach} and labeled as ''1p``.
\par
In this work, we focus on the electron spectrum in the \textit{unpolarized} case.
Therefore, the final step in both approaches involves summing over all angular quantum numbers of the final-state electron and averaging over projections of the initial-state muon's total angular momentum.
The resulting expression for the electron spectrum is given by
\begin{equation}
\begin{aligned}
\frac{dW^{1\mathrm{p},2\mathrm{p}}\left(E_{e},n_{\mu}\varkappa_{\mu}\right)}{dE_{e}} & =\frac{1}{\Pi_{j_{\mu}}^2}\sum_{\mu_{\mu}} \sum_{\varkappa_{e}\mu_{e}}\frac{dW^{1\mathrm{p},2\mathrm{p}}\left(\zeta_{e},\zeta_{\mu}\right)}{dE_{e}},
\end{aligned}
\end{equation}
where $\Pi_{a}$ is defined by Eq.~\eqref{eq:pi}.
\par
The two-particle approach is based on the multipole expansion of the two-particle interaction operator given in Eq.~\eqref{eq:fermi_interaction_operator}.
The corresponding multipole expansion, detailed in Appendix~\ref{app:a}, takes the form
\begin{equation}\label{eq:fermi_interaction_operator_multipole}
V^{\mathrm{F}}\left(1,2\right)=\frac{G_{\mathrm{F}}}{\sqrt{2}}\sum_{l}  \left[V_l^{\mathrm{s}}\left(1,2\right)-V_l^{\mathrm{v}}\left(1,2\right)\right],
\end{equation}
where $V_l^{\mathrm{s}}\left(1,2\right)$ and $V_l^{\mathrm{v}}\left(1,2\right)$ are the scalar and vector multipole components of the interaction, defined in Eqs.~\eqref{eq:fermi_interaction_operator_s_o} and~\eqref{eq:fermi_interaction_operator_v_o}, respectively.
Appendix~\ref{app:b} provides the derivation of the matrix elements of the operator in Eq.~\eqref{eq:fermi_interaction_operator_multipole} within the central-field approximation, while the summation over the total angular-momentum projections is presented in Appendix~\ref{app:c}.
The resulting expression for the electron spectrum is given by
\begin{widetext}
\begin{equation}\label{eq:e_spectrum_2body}
\frac{dW^{2\mathrm{p}}\left(E_{e},n_{\mu}\varkappa_{\mu}\right)}{dE_{e}}=\frac{G_{\mathrm{F}}^{2}}{16\pi}\sum_{\varkappa_{e}\varkappa_{\nu_{\mu}}\varkappa_{\nu_{e}}l}\left(\Pi_{lj_{\nu_{\mu}}}C_{l0j_{\mu}\frac{1}{2}}^{j_{e}\frac{1}{2}}C_{l0j_{\nu_{\mu}}\frac{1}{2}}^{j_{\nu_{e}}\frac{1}{2}}\right)^{2}\int_{0}^{E_{\mu}-E_{e}}dE_{\nu_{\mu}}\left|R_{l}^{\mathrm{s}}\left(e\nu_{\mu},\mu\nu_{e}\right)+\sum_{L=l-1}^{L=l+1}\frac{\Pi_{l}^{2}}{\Pi_{L}^{2}}R_{Ll}^{\mathrm{v}}\left(e\nu_{\mu},\mu\nu_{e}\right)\right|^{2},
\end{equation}
\end{widetext}
where the integration is performed over the energy of~$\nu_\mu$,~$E_\mu$ is the energy of the muon including its rest mass, and the energy of~$\nu_e$ is determined by the relation~$E_{\nu_e} = E_{\nu_{\mu}} - \left(E_\mu - E_e\right)$.
The other notations used in Eq.~\eqref{eq:e_spectrum_2body} are as follows: $C_{a\alpha b\beta}^{c\gamma}$ are the Clebsch-Gordan coefficients and the two-particle radial integrals~$R_{l}^{\mathrm{s}}\left(cd,ab\right)$ and~$R_{Ll}^{\mathrm{v}}\left(cd,ab\right)$ are defined in Eqs.~\eqref{eq:2p_rads_cdab} and~\eqref{eq:2p_radv_cdab}, respectively.
\par
The effective one-particle approach is based on a separation of lepton and neutrino degrees of freedom in the squared amplitude~$|A|^2$.
This separation can be accomplished by employing the explicit form of the massless neutrino wave functions, applying standard techniques to deal with the traces of $\gamma$ matrices, and introducing an effective lepton current defined as
\begin{equation}\label{eq:effective_current}
    J^{\alpha}\left(e\mu;\vec{k}\right)=\braket{e|\gamma^{0}\gamma^{\alpha}\left(1-\gamma^{5}\right)e^{-i\vec{k}\cdot\vec{r}}|\mu}.
\end{equation}
The details of the corresponding derivation are given in Appendix~\ref{app:d}.
The multipole expansion of the matrix elements of the effective current~$J^{\alpha}\left(e\mu;\vec{k}\right)$ within the central-field approximation is presented in Appendix~\ref{app:e}.
The integration over the polar and azimuthal angles of~$\vec{k}$ is carried out in Appendix~\ref{app:f}.
The resulting expression for the electron spectrum in the effective one-particle approach is
\begin{widetext}
\begin{equation}\label{eq:e_spectrum_1body}
\begin{aligned}
\frac{dW^{1\mathrm{p}}\left( E_{e},n_{\mu}\varkappa_{\mu}\right)}{dE_{e}}=\frac{G_{\mathrm{F}}^{2}}{12\pi^{3}}\frac{1}{\Pi_{j_{\mu}}^{2}}\sum_{\varkappa_{e}l}\left(\Pi_{j_{e}} C_{j_{e}\frac{1}{2}l0}^{j_{\mu}\frac{1}{2}}\right)^{2}\int_{0}^{E_{\mu}-E_{e}} & dk\,k^{2}\bigg\{ k^{2}R_{l}^{\mathrm{ss}}\left(e\mu;k\right)-2k\left(E_{\mu}-E_{e}\right)\mathrm{Re}R_{l}^{\mathrm{sv}}\left(e\mu;k\right)\\
 & +\left[\left(E_{\mu}-E_{e}\right)^{2}-k^{2}\right]R_{l}^{\mathrm{vv},1}\left(e\mu;k\right)+k^{2}R_{l}^{\mathrm{vv},2}\left(e\mu;k\right)\bigg\},
\end{aligned}
\end{equation}
\end{widetext}
where the effective one-particle radial integrals~$R_{l}^{x}\left(e\mu;k\right)$ correspond to the scalar-scalar (ss), scalar-vector (sv), vector-vector diagonal (vv,1), and vector-vector non-diagonal (vv,2) contributions.
These are defined in Eqs.~\eqref{eq:r_ss_oneparticle},~\eqref{eq:r_sv_oneparticle},~\eqref{eq:r_vv1_oneparticle}, and~\eqref{eq:r_vv2_oneparticle}, respectively.
\par
The expressions obtained in Eqs.~\eqref{eq:e_spectrum_2body} and~\eqref{eq:e_spectrum_1body} are mathematically equivalent.
While Eq.~\eqref{eq:e_spectrum_2body} possesses a more transparent analytical structure, Eq.~\eqref{eq:e_spectrum_1body} is better suited for numerical evaluation.
An additional advantage of Eq.~\eqref{eq:e_spectrum_2body} lies in its explicit inclusion of the neutrino wave functions, making it particularly useful for extending the analysis to scenarios which involve hypothetical interactions beyond the Standard Model.
A comparison between Eqs.~\eqref{eq:e_spectrum_2body} and~\eqref{eq:e_spectrum_1body} allows one to derive a definite-integral relation involving the spherical Bessel functions.
The derivation of this relation, along with a discussion of several special cases, is provided in Appendix~\ref{app:g}.

\section{Numerical results}\label{sec:3_results}
We present our results for the electron spectrum in the bound-muon decay process, normalized to the total decay rate of a free muon, $W_0$, as
\begin{equation}
    N\left(E_e\right) = \frac{1}{W_0}\frac{dW\left(E_e\right)}{dE_e},\qquad W_{0}=\frac{G_{\mathrm{F}}^{2}m_{\mu}^{5}}{192\pi^{3}},
\end{equation}
with~$m_{\mu}$ being the muon mass.
The structure of the expression for the electron spectrum reveals two principal sources of dependence on the details of the treatment of the muon and electron states.
The first one arises from the muon energy, which appears both in the upper integration limit and explicitly in the integrand.
The second one stems from the muon and electron wave functions, which enter the radial integrals.
\par
To isolate the effect of a specific \emph{corr}ection to the muon binding energy on the electron spectrum near its endpoint, we define the corresponding relative change as
\begin{equation}\label{eq:corr_enrg}
\delta^{\mathrm{corr}}_{\mathrm{enrg}} \left(E_e\right) = \left(\frac{E_{\mu} + \delta E_\mu^{\mathrm{corr}} - E_e}{E_{\mu} - E_e}\right)^5 - 1,
\end{equation}
where~$\delta E_\mu^{\mathrm{corr}}$ denotes the energy correction under consideration.
The characteristic power-law behavior of the spectrum near the endpoint can be derived by taking the limit~$E_e \to E_\mu$ in, e.g., Eq.~\eqref{eq:e_spectrum_1body}; see Refs.~\cite{1982_ShankerO_PRD25, 2011_CzarneckiA_PRD84} for a detailed discussion.
\par
To assess the sensitivity of the spectrum to a correction affecting the wave functions, we introduce the relative deviation defined by
\begin{equation}\label{eq:delta_corr_wf}\delta^{\mathrm{corr}}_{\mathrm{wf}} \left(E_e\right) = \frac{N^{\mathrm{corr}} \left(E_e +\delta E_\mu^{\mathrm{corr}} \right) - N\left(E_e \right)}{N\left(E_e \right)},
\end{equation}
where~$N^{\mathrm{corr}} \left(E_e +\delta E_\mu^{\mathrm{corr}} \right)$ is the spectrum computed using the corrected wave functions and for the electron energy shifted by the corresponding correction to the muon binding energy, $\delta E_\mu^{\mathrm{corr}}$.
We note that~$E_\mu$ and~$E_e$ enter into the computational formulas in the combination~$E_\mu-E_e$, therefore the argument shift in Eq.~\eqref{eq:delta_corr_wf} serves to neutralize the correction to~$E_\mu$.
Finally, the total relative deviation of the spectrum due to a given correction is defined as
\begin{equation}\label{eq:delta_corr_tot}
    \delta^{\mathrm{corr}} \left(E_e\right)= \frac{N^{\mathrm{corr}} \left(E_e \right) - N\left(E_e \right)}{N\left(E_e \right)}.
\end{equation}
\par
The radial wave functions for the bound muon and the unbound electron are obtained by numerically solving the radial Dirac equation on a discrete grid using the package RADIAL~\cite{1995_SalvatF_CPC90, *2019_SalvatF_CPC240}.
The radial integrals are evaluated employing a standard Gauss-Legendre quadrature scheme, adapted to the spatial region in which the muon radial wave function remains numerically significant.
The integration over the energy is performed using the same quadrature method.
In addition to the central Coulomb potential, several atomic and quantum electrodynamical (QED) corrections  are incorporated for both the muon and the electron.
These include the finite-nuclear-size (FNS), nuclear-deformation (ND), vacuum-polarization (VP), and electron-screening (SCR) effects.
Each correction is taken into account by means of an appropriate local potential inserted directly into the radial Dirac equation, thereby influencing both the wave functions and the muon binding energy.
We also consider two distinct recoil (REC) corrections.
\par
The FNS effect is treated using the Fermi nuclear-charge distribution model, with the parameters adopted from Ref.~\cite{Ang13}.
The ND effect is incorporated following the approach of Ref.~\cite{2008_KozhedubY_PRA77}, in which the ND potential is constructed numerically by averaging a modified Fermi nuclear potential over angular coordinates.
The latter treatment employs the standard $\beta$-parametrization of ND.
The ND parameters~$\beta_{20}$ for the even-even nuclei~${}^{12}_{\,\,\,6}\mathrm{C}$ and~${}^{28}_{14}\mathrm{Si}$ are taken from Ref.~\cite{2016_PritychenkoB_ADNDT107}.
For the odd-proton even-neutron nucleus~${}^{27}_{13}\mathrm{Al}$, the ND effect is not included.
\par
As for the VP effect, we consider two contributions: the leading-order one due to the Uehling (Ue) potential and a correction due to the Wichmann-Kroll (WK) potential.
These local VP potentials are generated using the QEDMOD package~\cite{2013_ShabaevV_PhysRevA, 2015_ShabaevV_CompPhysComm, *2018_ShabaevV_CompPhysComm}, which implements methods for the Ue potential described in Ref.~\cite{1976_FullertonL_PRA13} and the approximate formulas for the WK potential derived in Ref.~\cite{1991_FainshteinA_JPB24}.
\par
The interaction between bound atomic electrons and the muon as well as the final-state unbound electron, is modeled by a local electron-screening (SCR) potential.
The first REC correction accounts for the shift in the binding energy of the initial muon induced by the nuclear recoil.
The leading-order-recoil contribution to this effect can be included using the non-relativistic mass-shift (MS) operator,~$\vec{p}^2/\left(2M_{\mathrm{n}}\right)$, where $M_{\mathrm{n}}$ is the nuclear mass.
This correction is evaluated as the expectation value of the MS operator with the FNS wave function.
According to Ref.~\cite{2023_YerokhinV_PRA108}, the deviation between this non-relativistic approach and a full QED treatment of the recoil correction does not exceed $2.5\%$ for the ground state of muonic hydrogen-like ions with $10\leq Z\leq 20$.
The second REC correction accounts for the recoil effect on the kinematics of the final-state electron following the procedure outlined in Ref.~\cite{2011_CzarneckiA_PRD84}; see also Refs.~\cite{1979_vonBaeyerH_PRA19, 1982_ShankerO_PRD25} for related discussions.
This correction is incorporated by replacing~$E_e\to E_e - E_e^2/\left(2M_n\right)$ in the expression for the electron spectrum, Eq.~\eqref{eq:e_spectrum_1body} as well as in Eq.~\eqref{eq:corr_enrg}.
This approximation is valid near the end point of the electron spectrum, where the momentum transfer to the neutrinos is minimal and the electron can be treated as a highly relativistic and effectively massless particle.
Within this approximation, the atomic mass is also replaced with the nuclear mass~$M_n$.
Nuclear masses are taken from the compilation in Ref.~\cite{2021_WangM_ChinPhysC45}.
\par
The present calculations are performed for the bound muon occupying the ground state characterized by~$n_\mu=1$ and~$\varkappa_\mu = -1$.
We consider three nuclear isotopes which are of experimental relevance:~${}^{12}_{\,\,\,6}\mathrm{C}$,~${}^{27}_{13}\mathrm{Al}$, and~${}^{28}_{14}\mathrm{Si}$.
One of our primary goals is the evaluating of the corrections to the muon binding energy,~$ E^\mathrm{bind}_{\mu}$, which arise from various atomic and nuclear effects discussed above.
The computed corrections for each isotope are summarized in Table~\ref{tab:1}, the uncertainties are given in parentheses.
The uncertainties of the FNS values are due to the uncertainties of tabulated root-mean-square nuclear-charge radii of the isotopes.
\begin{table}[htbp]
\centering
\caption{
Contributions to the ground-state binding energy of the muon,~$E^\mathrm{bind}_{\mu} = E_\mu - m_\mu c^2$, in selected muonic hydrogen-like ions,~in a.u..
The first row,~$E^{\mathrm{Dirac}}$, shows the Dirac energy of the muon, bounded by the Coulomb potential of a point-like nucleus, with the rest mass subtracted.
The second row,~$\delta E^{\mathrm{FNS}}$, presents the correction to the Dirac energy resulting from the inclusion of finite-nuclear-size (FNS) effect.
The uncertainties in the parentheses show the errors associated with the uncertainties of the root-mean-square radii.
Subsequent rows list corrections to~$E^{\mathrm{Dirac}}+\delta E^\mathrm{FNS}$ due to the nuclear-deformation (ND), mass-shift (MS), Uehling (Ue), Wichmann-Kroll (WK), and electron-screening (SCR) effects.
The uncertainties of the MS corrections account for the omitted QED contributions, while the SCR uncertainties arise from an analysis based on using different screening potentials.
The final row provides the total energy of the muon,~$E_{\mu}$, including its rest mass.
}
\begin{tabular}{
l
S[table-format=-4.4(1)]
S[table-format=-4.4(1)]
S[table-format=-4.4(2)]
}
\toprule
\multicolumn{1}{c}{} & \multicolumn{1}{c}{${}^{12}_6\mathrm{C}$} & \multicolumn{1}{c}{${}^{27}_{13}\mathrm{Al}$} & \multicolumn{1}{c}{${}^{28}_{14}\mathrm{Si}$}\\

\midrule
$E^{\mathrm{Dirac}}$ & -3723.61 & -17511.4 & -20316.4\\
$\delta E^\mathrm{FNS}$ &15.03(3) &430.0(8)  &586.2(8)       \\
$\delta E^\mathrm{ND}$    &0.01      &         &-0.4       \\
$\delta E^\mathrm{MS}$   &34.90(5)     & 69.6(9)   &\multicolumn{1}{S[table-format=-4.4(2),parse-numbers=false]}{77.1(1.2)}     \\
$\delta E^\mathrm{Ue}$     &-14.81    & -98.7   &-117.1     \\
$\delta E^\mathrm{WK}$    &0.002     & 0.04    &0.05       \\
$\delta E^\mathrm{SCR}$   &8(3)  & 33(9)    &35(7)       \\
\midrule
$ E^\mathrm{bind}_{\mu}$   &-3681(3)  & -17078(9) &-19736(7)       \\
$E^\mathrm{bind}_{\mu}\mathrm{\,\,\,[MeV]}$     & -0.10015(8)  & -0.4647(2)      & -0.5370(2)      \\
\midrule
$E_{\mu}\mathrm{\,\,\,\,\,\,\,\,\,\,[MeV]}$     & 105.55822(8)  & 105.1937(2)      & 105.1213(2)      \\
\bottomrule
\end{tabular}
\label{tab:1}
\end{table}
\par
For~$Z=6$, the dominant contribution to the muon binding energy~$E^\mathrm{bind}_{\mu}$ arises from the MS correction, $\delta E^\mathrm{MS}$,
while the FNS correction, $\delta E^\mathrm{FNS}$, and the Ue correction, $\delta E^\mathrm{Ue}$, are more than two times smaller and almost cancel each other out, having the opposite signs.
However, for~$Z=13$ and $Z=14$, the FNS correction considerably overweights the Ue and MS ones.
The uncertainty associated with the MS correction is estimated based on omitted QED contributions to the nuclear recoil effect using the tabulated values from Ref.~\cite{2023_YerokhinV_PRA108}.
It is worth noting that the Ue and MS corrections have the opposite signs and partly cancel each other.
To evaluate the SCR correction, we consider three screening potentials from the $X\alpha$-family choosing $X\alpha=0$,~$2/3$, and~$1$; for details see e.g. Ref.~\cite{2002_SapirsteinJ_PRA66}.
To partially take into account the reconfiguration of the electron shells induced by the presence of the muon, the so-called~$Z-1$ approximation~\cite{1982_BorieE_RMP54} is used.
Specifically, the screening potentials are generated for~$Z=5$ with the electronic configuration~$1s^22s^2$ in the case of carbon, and for~$Z=12$ and~$Z=13$ with the configuration~$1s^22s^22p^63s^2$ in case of aluminum and silicon, respectively.
The correction to the muon binding energy due to the electron-screening effects is taken as the average of the results obtained for these three potentials
The corresponding uncertainties are estimated conservatively as standard deviations of these values.
The electron-screening correction to the muon binding energy is found to be significant: approximately~$8(3)$~a.u. for~$Z=6$,~$33(9)$~a.u. for~$Z=13$, and~$35(7)$~a.u. for~$Z=14$.
We note that including additional electrons in the configurations used to determine the screening potentials yields the SCR corrections that are within the estimated uncertainties.
In contrast, both WK and ND corrections to the ground-state energy are found to be small for all the considered values of~$Z$.
Nevertheless, as it will be demonstrated in the subsequent analysis, the total correction to the muon binding energy solely does not fully determine the behavior of the electron spectrum near its endpoint.
The influence of the corrections to the wave functions must also be taken into account to achieve a complete description.
\par
We compare our calculated muon ground-state energies with available literature data to validate the numerical approach.
For aluminum, the energy including the FNS effect,~$E^{\mathrm{FNS}}=E^{\mathrm{Dirac}} + \delta E^{\mathrm{FNS}}$, is $-0.46481(2)$~MeV which is in reasonable agreement with the result reported in Ref.~\cite{2011_CzarneckiA_PRD84},~$E^{\mathrm{FNS}}=-0.464$~MeV.
Our computed Ue correction to the muon binding energy for~${}^{27}_{13}\mathrm{Al}$,~$\delta E^{\mathrm{Ue}}=-98.7$ a.u., is consistent with the value from ~\cite{2016_SzafronR_PRD94},~$\delta E^{\mathrm{Ue}}=-99$~a.u..
Furthermore, the correction due to the electron screening for~$Z=13$,~$\delta E^{\mathrm{SCR}}=33(9)$~a.u., agrees with the estimate given in Ref.~\cite{2011_CzarneckiA_PRD84},~$\delta E^{\mathrm{SCR}}=40$~a.u..
These comparisons demonstrate good overall agreement with previously published results and provide additional confidence in the reliability of the employed numerical framework.
\par
We now investigate the impact of the FNS, ND, Ue, WK, and SCR corrections on the electron spectrum near the endpoint.
Our analysis focuses on the isotopes~${}^{12}_{\,\,\,6}\mathrm{C}$ and~${}^{28}_{14}\mathrm{Si}$.
Since the nuclear charge of silicon differs from that of aluminum by only one unit, qualitative differences in the electron spectrum of these two systems are hardly observed.
Throughout the following discussion, the kinematic REC correction to the final-state electron is consistently included.
\par
\begin{figure}[htpb]
\centering
\includegraphics[width=\columnwidth]{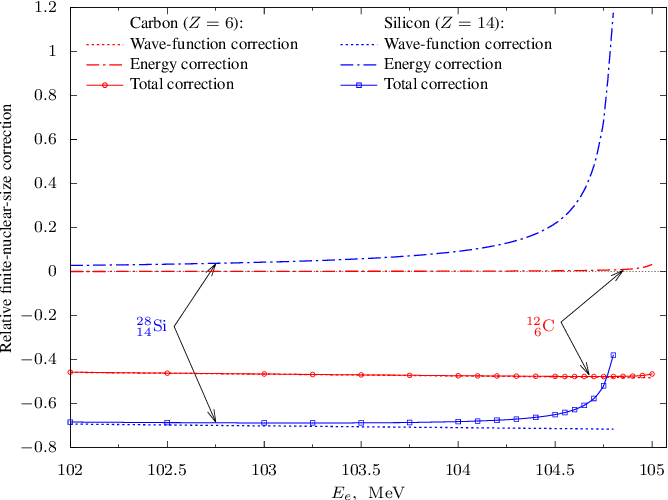}
\caption{
Relative finite-nuclear-size corrections to the electron spectrum near the endpoint in the ground-state bound-muon decay process for~${}^{12}_{\,\,\,6}\mathrm{C}$ and~${}^{28}_{14}\mathrm{Si}$ nuclei.
Dotted, dash-dotted, and solid lines correspond to the wave-function, energy, and total corrections, respectively.}
\label{fig:fns_relative_correction}
\end{figure}

We begin by examining the dominant correction to the muon bound-state energy, which arises from the FNS effects.
In Fig.~\ref{fig:fns_relative_correction}, we present the relative FNS-induced corrections to the electron spectrum near the endpoint: the energy correction~$\delta^{\mathrm{FNS}}_{\mathrm{enrg}}\left(E_e\right)$, the wave-function correction~$\delta^{\mathrm{FNS}}_{\mathrm{wf}}\left(E_e\right)$, and the total correction~$\delta^{\mathrm{FNS}}\left(E_e\right)$ defined in Eqs.~\eqref{eq:corr_enrg}, \eqref{eq:delta_corr_wf}, and~\eqref{eq:delta_corr_tot}, respectively.
For~$Z=6$, the FNS effect is almost completely due to the wave-function correction and is about of~$-44\%$ close to the endpoint.
In the case of~$Z=14$, this correction is even more pronounced, constituting~$-68\%$.
Notably, for~$Z=6$,~$\delta^{\mathrm{FNS}}_{\mathrm{wf}}\left(E_e\right)$ outweighs the corresponding energy correction by several orders of magnitude.
It should be noted that in the previous study~\cite{1993_WatanabeR_ADNDT54}, the point-like nucleus was assumed for~$^{16}_{\,\,\,8}\mathrm{O}$.
However, our results indicate that even for lower values of~$Z$, the FNS effects can significantly alter the spectrum near the endpoint.
For~$Z=14$, while the energy correction becomes more substantial than the wave-function one, the total correction to the spectrum still remains governed by the latter: even near the very spectrum endpoint the sign of the total FNS correction
coincide with the sign for the wave-function one.
This is due to the approximate nature of the separation~$\delta^{\mathrm{FNS}}\left(E_e\right) \approx \delta^{\mathrm{FNS}}_{\mathrm{enrg}}\left(E_e\right) + \delta^{\mathrm{FNS}}_{\mathrm{wf}}\left(E_e\right)$, which becomes less accurate as higher-order FNS effects grow with increasing~$Z$.
Furthermore, in both isotopes considered,~$\delta^{\mathrm{FNS}}_{\mathrm{wf}}\left(E_e\right)$ exhibits a common linear decrease with increasing~$E_e$ near the endpoint.
We conclude that the FNS effect must be treated self-consistently by incorporating its influence on the wave functions.
\par
\begin{figure}[htbp]
\centering
\includegraphics[width=\columnwidth]{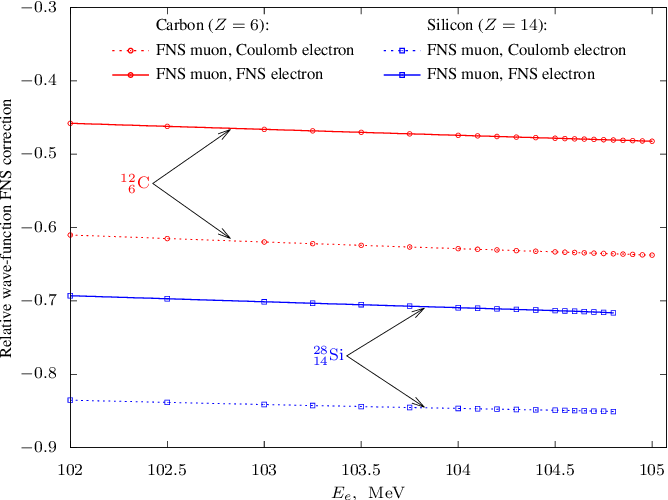}
\caption{
Relative wave-function correction due to finite-nuclear-size (FNS) effects on the electron spectrum near the endpoint in the bound-muon decay process for~${}^{12}_{\,\,\,6} \mathrm{C}$ and ${}^{28}_{14} \mathrm{Si}$ nuclei.
The dotted lines represents the case where the emitted electron is subjected to the Coulomb potential of a point-like nucleus, while the solid lines corresponds to the case where the FNS effect is included in the nuclear potential.}
\label{fig:fns_relative_correction_electron}
\end{figure}

To determine whether the FNS effect are relevant only for the muon or also for the unbound electron, we analyze the relative wave-function correction to the spectrum under different nuclear potentials for the final-state electron.
Specifically, we consider two scenarios: (i) the electron is subjected to the pure Coulomb potential of a point-like nucleus; and (ii) the electron is subjected to the Fermi-distributed nuclear potential.
The corresponding results are shown in Fig.~\ref{fig:fns_relative_correction_electron}.
We observe that the difference in~$\delta^{\mathrm{FNS}}_{\mathrm{wf}}\left(E_e\right)$ between the cases (i) and (ii) is approximately~$-15\%$ for both~$Z=6$ and~$Z=14$.
This notable discrepancy demonstrates that, even for low-$Z$ nuclei, the FNS effects on the electron must be treated self-consistently to ensure accurate predictions for the spectrum.
\par
\begin{figure}[htbp]
\centering
\includegraphics[width=\columnwidth]{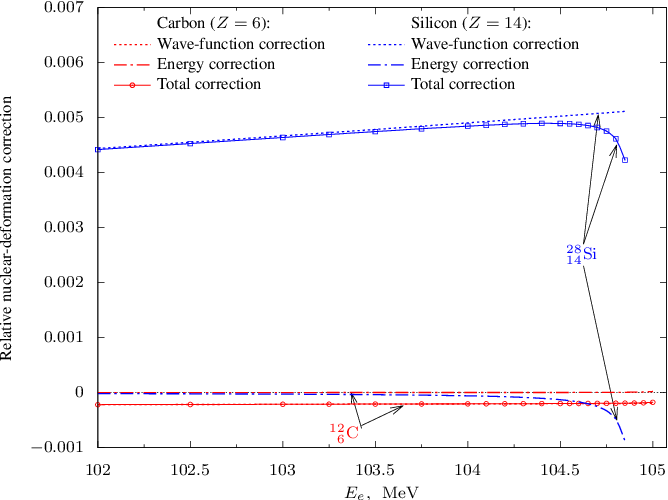}
\caption{Relative nuclear-deformation corrections to the electron spectrum near the endpoint in the ground-state bound-muon decay process for~${}^{12}_{\,\,\,6}\mathrm{C}$ and~${}^{28}_{14}\mathrm{Si}$ nuclei.
Dotted, dash-dotted, and solid lines correspond to the wave-function, energy, and total corrections, respectively.}
\label{fig:nd_relative_correction}
\end{figure}

We now turn our attention to the ND effects.
Since ND corrections represent a small perturbation to the standard Fermi nuclear charge distribution model, we expect their influence on the spectrum to exhibit a similar pattern to that observed for the FNS corrections.
This expectation is confirmed by the results presented in Fig.~\ref{fig:nd_relative_correction}, which show the relative ND corrections to the spectrum.
Near the endpoint, for both isotopes under consideration, the ND correction is entirely governed by the modified wave functions, while the corresponding energy correction to the muon ground state is several orders of magnitude smaller and thus negligible.
For~$Z=6$, the wave-function correction is small, approximately~$-0.02\%$, whereas for~$Z=14$, it not only changes the sign but also becomes significantly larger, reaching about~$0.4\%$.
These findings clearly indicate that the ND effect must be incorporated directly into the wave functions.
Simply accounting for the ND correction to the muon binding energy is insufficient to account for the ND correction on the electron spectrum.
\par
\begin{figure}[htbp]
\centering
\includegraphics[width=\columnwidth]{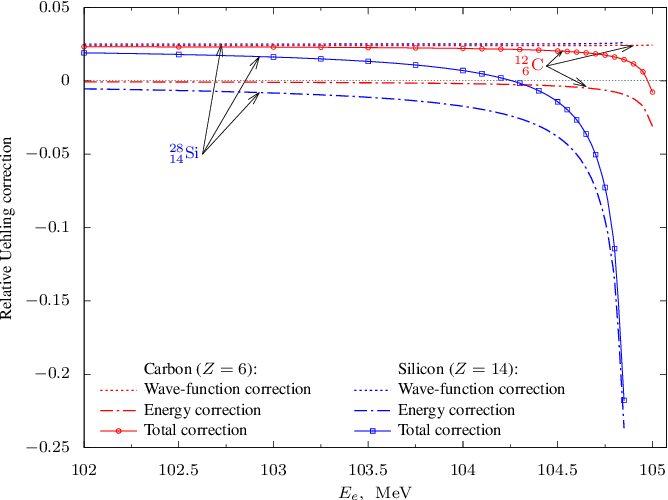}
\caption{Relative Uehling corrections to the electron spectrum near the endpoint in the ground-state bound-muon decay process for~${}^{12}_{\,\,\,6}\mathrm{C}$ and~${}^{28}_{14}\mathrm{Si}$ nuclei.
Dotted, dash-dotted, and solid lines correspond to the wave-function, energy, and total corrections, respectively.}
\label{fig:ue_relative_correction}
\end{figure}

In Fig.~\ref{fig:ue_relative_correction}, we present the relative Uehling corrections to the electron spectrum near the endpoint in the bound-muon decay process, including the energy correction~$\delta^{\mathrm{Ue}}_{\mathrm{enrg}}\left(E_e\right)$, the wave-function correction~$\delta^{\mathrm{Ue}}_{\mathrm{wf}}\left(E_e\right)$, and the total correction~$\delta^{\mathrm{Ue}}\left(E_e\right)$.
Although the Ue correction to the muon binding energy for~$Z=14$ is approximately ten times larger than that for~$Z=6$, the corresponding wave-function correction~$\delta^{\mathrm{Ue}}_{\mathrm{wf}}\left(E_e\right)$ exhibits minimal dependence on the electron energy and remains nearly constant at about~$2.2\%$ for both nuclei.
In contrast, the energy correction~$\delta^{\mathrm{Ue}}_{\mathrm{enrg}}\left(E_e\right)$ becomes significant only in the vicinity of the endpoint.
Our calculated values for the relative Ue corrections to the spectrum, $\delta^{\mathrm{Ue}}_{\mathrm{wf}}\left(E_e\right)$, $\delta^{\mathrm{Ue}}_{\mathrm{enrg}}\left(E_e\right)$, and $\delta^{\mathrm{Ue}}_{\mathrm{tot}}\left(E_e\right)$, in the case of~${}^{27}_{13}\mathrm{Al}$ are consistent with the findings reported in Ref.~\cite{2016_SzafronR_PRD94}.
\par
\begin{figure}[htbp]
\centering
\includegraphics[width=\columnwidth]{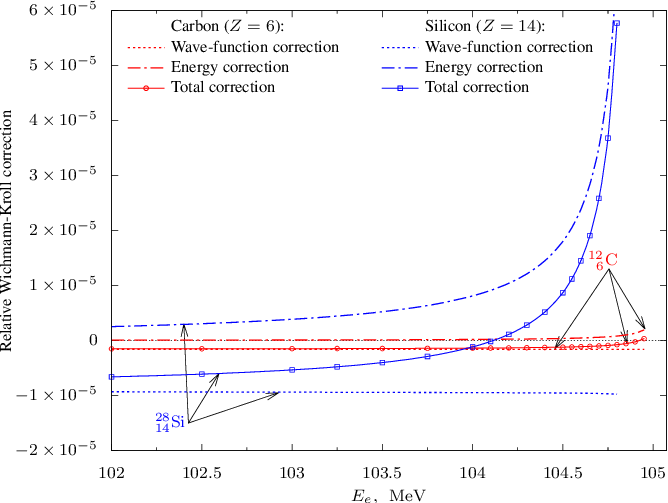}
\caption{
Relative Wichmann–Kroll corrections to the electron spectrum near the endpoint in the ground-state bound-muon decay process for~${}^{12}_{\,\,\,6}\mathrm{C}$ and~${}^{28}_{14}\mathrm{Si}$ nuclei.
Dotted, dash-dotted, and solid lines correspond to the wave-function, energy, and total corrections, respectively.}
\label{fig:wk_relative_correction}
\end{figure}

The WK corrections, shown in Fig.~\ref{fig:wk_relative_correction}, display a different behavior compared to the Ue case.
Notably, the wave-function correction~$\delta^{\mathrm{WK}}_{\mathrm{wf}}\left(E_e\right)$ is no longer similar for ~$Z=6$ and~$Z=14$.
The corresponding wave-function correction~$\delta^{\mathrm{WK}}_{\mathrm{wf}}\left(E_e\right)$ is about four times larger for~$Z=14$ than for~$Z=6$.
Nevertheless, the absolute magnitude of this correction remains negligible, about~$0.0002\%$ for~$Z=6$, which is four orders of magnitude smaller than the Ue correction~$\delta^{\mathrm{Ue}}\left(E_e\right)$, and about~$0.005\%$ for~$Z=14$.
Therefore, the WK contribution to the spectrum can be safely neglected in practical calculations.
\par
\begin{figure}[htbp]
\centering
\includegraphics[width=\columnwidth]{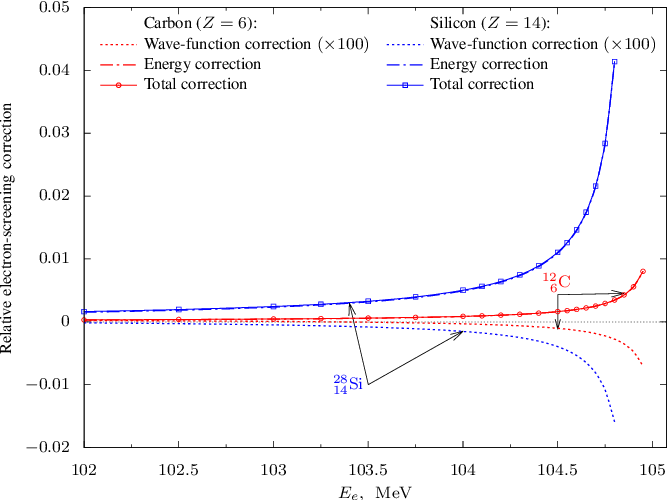}
\caption{Relative electron-screening corrections to the electron spectrum near the endpoint in the ground-state bound-muon decay process for~${}^{12}_{\,\,\,6}\mathrm{C}$,~${}^{27}_{13}\mathrm{Al}$, and~${}^{28}_{14}\mathrm{Si}$ nuclei.
The dash-dotted lines represents the energy correction, while the dotted lines show the wave-function correction, magnified by a factor of~$100$ for visibility.
The solid lines correspond to the total corrections which closely coincides with the energy corrections.}
\label{fig:scr_relative_correction}
\end{figure}

We turn to the final correction affecting wave functions: the electron-screening effect.
The corresponding results are shown in Fig.~\ref{fig:scr_relative_correction}.
In contrast to the FNS and ND corrections, the SCR effect exhibits a completely opposite trend.
Specifically, the correction is fully determined by the energy shift, while the wave-function contribution is negligible: less than~$-0.01\%$ for both~${}^{12}_{\,\,\,6}\mathrm{C}$ and~${}^{28}_{14}\mathrm{Si}$.
This indicates that, for all considered systems, the electron-screening effect near the endpoint does not require a self-consistent treatment for the wave functions.
Instead, it is sufficient to account for the SCR effect solely through a correction to the endpoint energy.
\par
Lastly, in Fig. 7, we present the total relative corrections to the electron spectrum near the endpoints for~${}^{12}_{\,\,\,6}\mathrm{C}$,~${}^{27}_{13}\mathrm{Al}$, and~${}^{28}_{14}\mathrm{Si}$.
\begin{figure}[htbp]
\centering
\includegraphics[width=\columnwidth]{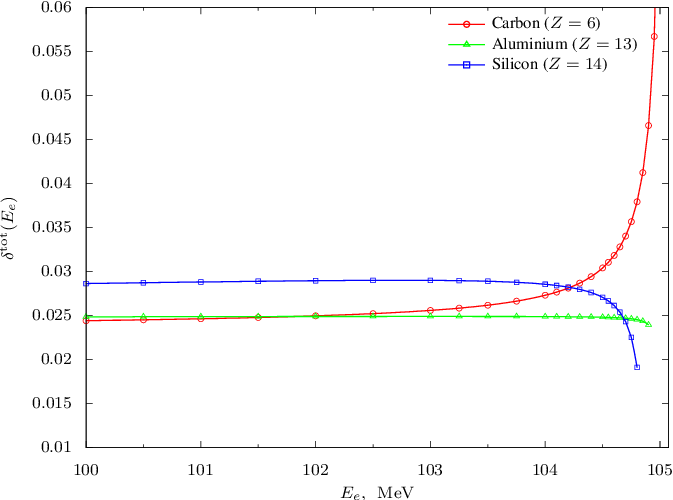}
\caption{Total relative corrections to the electron spectrum near the endpoints in the bound-muon decay process in~${}^{12}_{\,\,\,6}\mathrm{C}$,~${}^{27}_{13}\mathrm{Al}$, and~${}^{28}_{14}\mathrm{Si}$ nuclei, incorporating simultaneously the finite-nuclear-size, Uehling, Wichmann–Kroll, nuclear deformation (ND), nuclear recoil, and electron-screening effects.
The ND correction is not applied for aluminum.}
\label{fig:total_relative_correction}
\end{figure}
For carbon, the total correction reaches approximately~$2.5\%$ in the energy range~$100\leq E_e\leq102.5$ MeV, where the wave-function corrections dominate.
The total correction then rapidly increases up to about~$4\%$ near~$E_e\approx104.5$ MeV, where the energy corrections become more significant.
In the case of silicon, the total relative correction is slightly larger, about~$2.8\%$, and remains nearly constant up to~$E_e\approx104$ MeV, after which it gradually decreases to~$2\%$ as the energy corrections begin to dominate.
Aluminum exhibits a distinct behavior: the total relative correction remains nearly constant at approximately~$2.5\%$ level until the very endpoint of the spectrum.
Indeed, this behavior can be anticipated from the beginning of our analysis.
Returning to Table~\ref{tab:1}, one can notice that the various corrections to the muon ground-state energy nearly cancel each other out, leading to a total total energy correction of only~$3(9)$ a.u..
As a result, for~$Z=13$, the total correction is almost entirely determined by the wave-function modifications.
\par
\begin{figure}[htbp]
\centering
\includegraphics[width=\columnwidth]{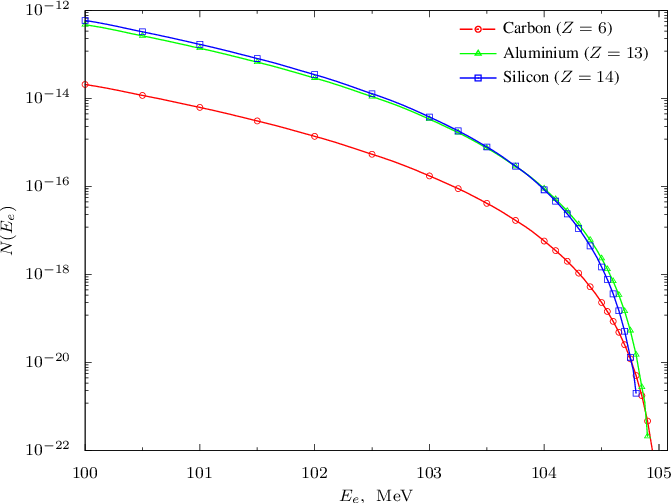}
\caption{Electron spectra near the endpoints in the bound-muon decay process for the~${}^{12}_{\,\,\,6}\mathrm{C}$,~${}^{27}_{13}\mathrm{Al}$, and~${}^{28}_{14}\mathrm{Si}$ nuclei, calculated with the inclusion of the finite-nuclear-size, Uehling, Wichmann–Kroll, nuclear-deformation (ND), nuclear-recoil, and electron-screening corrections.
The ND correction is not applied for aluminum.}
\label{fig:e_spectra_final}
\end{figure}

Finally, our analysis is completed by Fig.~\ref{fig:e_spectra_final}, which shows the electron spectra for~${}^{12}_{\,\,\,6}\mathrm{C}$,~${}^{27}_{13}\mathrm{Al}$, and~${}^{28}_{14}\mathrm{Si}$, including the FNS, Ue, WK, ND, NR, and SCR effects.

\section{Conclusion}\label{sec:4_conclusion}
In this study of the bound-muon decay process, we investigated within the framework of the Fermi’s effective theory the influence of various atomic effects on the electron spectrum near its endpoint.
We derived two equivalent expressions for the electron spectrum corresponding to an arbitrary initial bound-muon state and demonstrated the connection between them, which led to the definite-integral relation involving products of the spherical Bessel functions.
We systematically analyzed the influence the of finite-nuclear-size, nuclear-deformation, electron-screening, and vacuum-polarization effects.
These corrections were treated self-consistently by incorporating them to both the bound muon and unbound electron wave functions, as well as to the muon binding energy.
The nuclear-recoil corrections were also included: for the initial-state muon via the perturbative approach using the non-relativistic mass-shift operator, and for the final electron by means of the kinematic approach.
The nuclear-deformation correction was implemented following the method of Ref.~\cite{2008_KozhedubY_PRA77}, while the electron screening was modeled using the local spherically symmetric screening potentials.
\par
Our analysis focused on the nuclear isotopes~${}^{12}_{\,\,\,6}\mathrm{C}$,~${}^{27}_{13}\mathrm{Al}$, and~${}^{28}_{14}\mathrm{Si}$.
We found that the finite-nuclear-size (FNS), nuclear-deformation, and vacuum-polarization corrections are mainly determined by self-consistent modifications of the wave functions.
Even for small nuclear-charge numbers, such as~$Z=6$, the FNS effect must be incorporated into the Dirac equation for both muon and electron.
In particular, for carbon, the FNS correction to the spectrum near the endpoint can exceed~$-40\%$, highlighting the inadequacy of the point-like nuclear model even for small~$Z$, though it was previously used for~$Z=8$ in Ref.~\cite{1993_WatanabeR_ADNDT54}.
The relative nuclear-deformation correction to the endpoint of the spectrum is significant for silicon~$(0.5\%)$ but negligible for carbon~$(-0.02\%)$.
In contrast, the SCR correction is entirely determined by its effect on the muon ground-state energy, with the wave-function contributions found to be negligible for all the cases considered.
\par
Taking into account all the effects, the total relative correction to the spectrum near the endpoint is approximately~$2.5\%$ for aluminum,~$2.8\%$ for silicon, and up to~$5\%$ for carbon.
These results provide an important supplement for upcoming experiments~\cite{2019_TeshimaN_arXiv1911, 2022_MoritsuM_MDPI8, 2025_MiscettiS_NIMPA1073} aimed at improving sensitivity to the charged lepton flavor-violating processes, such as the neutrino-less bound-muon-to-electron conversion.

\section{Acknowledgments}
\label{sec:5_acknowledgments}
Research of M.~Y.~K. and A.~O.~D. was supported by Natural Sciences and Engineering Research Canada (NSERC).
The work of A. V. M. and Y. S. K. was supported by the Russian Science Foundation (Grant No. 22-62-00004, https://rscf.ru/project/22-62-00004/).

\bibliographystyle{apsrev}
\bibliography{main}

\begin{thebibliography}{44}
\expandafter\ifx\csname natexlab\endcsname\relax\def\natexlab#1{#1}\fi
\expandafter\ifx\csname bibnamefont\endcsname\relax
  \def\bibnamefont#1{#1}\fi
\expandafter\ifx\csname bibfnamefont\endcsname\relax
  \def\bibfnamefont#1{#1}\fi
\expandafter\ifx\csname citenamefont\endcsname\relax
  \def\citenamefont#1{#1}\fi
\expandafter\ifx\csname url\endcsname\relax
  \def\url#1{\texttt{#1}}\fi
\expandafter\ifx\csname urlprefix\endcsname\relax\def\urlprefix{URL }\fi
\providecommand{\bibinfo}[2]{#2}
\providecommand{\eprint}[2][]{\url{#2}}

\bibitem[{\citenamefont{Marciano et~al.}(2008)\citenamefont{Marciano, Mori, and Roney}}]{2008_MarcianoW_AnnRevNuclPartPhys58}
\bibinfo{author}{\bibfnamefont{W.~J.} \bibnamefont{Marciano}}, \bibinfo{author}{\bibfnamefont{T.}~\bibnamefont{Mori}}, \bibnamefont{and} \bibinfo{author}{\bibfnamefont{J.~M.} \bibnamefont{Roney}}, \bibinfo{journal}{Annual Review of Nuclear and Particle Science} \textbf{\bibinfo{volume}{58}}, \bibinfo{pages}{315} (\bibinfo{year}{2008}).

\bibitem[{\citenamefont{Bernstein and Cooper}(2013)}]{2013_BernsteinR_PR532}
\bibinfo{author}{\bibfnamefont{R.~H.} \bibnamefont{Bernstein}} \bibnamefont{and} \bibinfo{author}{\bibfnamefont{P.~S.} \bibnamefont{Cooper}}, \bibinfo{journal}{Physics Reports} \textbf{\bibinfo{volume}{532}}, \bibinfo{pages}{27} (\bibinfo{year}{2013}).

\bibitem[{\citenamefont{Calibbi and Signorelli}(2018)}]{2018_CalibbiL_NouvCim41}
\bibinfo{author}{\bibfnamefont{L.}~\bibnamefont{Calibbi}} \bibnamefont{and} \bibinfo{author}{\bibfnamefont{G.}~\bibnamefont{Signorelli}}, \bibinfo{journal}{La Rivista del Nuovo Cimento} \textbf{\bibinfo{volume}{41}}, \bibinfo{pages}{71} (\bibinfo{year}{2018}).

\bibitem[{\citenamefont{Ardu and Pezzullo}(2022)}]{2022_ArduM_MDPI8}
\bibinfo{author}{\bibfnamefont{M.}~\bibnamefont{Ardu}} \bibnamefont{and} \bibinfo{author}{\bibfnamefont{G.}~\bibnamefont{Pezzullo}}, \bibinfo{journal}{Universe} \textbf{\bibinfo{volume}{8}}, \bibinfo{pages}{299} (\bibinfo{year}{2022}).

\bibitem[{\citenamefont{Bertl et~al.}(2006)\citenamefont{Bertl, Engfer, Hermes, Kurz, Kozlowski, Kuth, Otter, Rosenbaum, Ryskulov, van~der Schaaf et~al.}}]{2006_BertlW_EPJC47}
\bibinfo{author}{\bibfnamefont{W.}~\bibnamefont{Bertl}}, \bibinfo{author}{\bibfnamefont{R.}~\bibnamefont{Engfer}}, \bibinfo{author}{\bibfnamefont{E.}~\bibnamefont{Hermes}}, \bibinfo{author}{\bibfnamefont{G.}~\bibnamefont{Kurz}}, \bibinfo{author}{\bibfnamefont{T.}~\bibnamefont{Kozlowski}}, \bibinfo{author}{\bibfnamefont{J.}~\bibnamefont{Kuth}}, \bibinfo{author}{\bibfnamefont{G.}~\bibnamefont{Otter}}, \bibinfo{author}{\bibfnamefont{F.}~\bibnamefont{Rosenbaum}}, \bibinfo{author}{\bibfnamefont{N.}~\bibnamefont{Ryskulov}}, \bibinfo{author}{\bibfnamefont{A.}~\bibnamefont{van~der Schaaf}}, \bibnamefont{et~al.}, \bibinfo{journal}{The European Physical Journal C - Particles and Fields} \textbf{\bibinfo{volume}{47}}, \bibinfo{pages}{337} (\bibinfo{year}{2006}).

\bibitem[{\citenamefont{Teshima}()}]{2019_TeshimaN_arXiv1911}
\bibinfo{author}{\bibfnamefont{N.}~\bibnamefont{Teshima}}, \bibinfo{note}{arXiv:1911.07143 [physics]}.

\bibitem[{\citenamefont{Moritsu}(2022)}]{2022_MoritsuM_MDPI8}
\bibinfo{author}{\bibfnamefont{M.}~\bibnamefont{Moritsu}}, \bibinfo{journal}{Universe} \textbf{\bibinfo{volume}{8}}, \bibinfo{pages}{196} (\bibinfo{year}{2022}).

\bibitem[{\citenamefont{Miscetti}(2025)}]{2025_MiscettiS_NIMPA1073}
\bibinfo{author}{\bibfnamefont{S.}~\bibnamefont{Miscetti}}, \bibinfo{journal}{Nuclear Instruments and Methods in Physics Research Section A: Accelerators, Spectrometers, Detectors and Associated Equipment} \textbf{\bibinfo{volume}{1073}}, \bibinfo{pages}{170257} (\bibinfo{year}{2025}).

\bibitem[{\citenamefont{Dzhilkibaev and Lobashev}(1989)}]{1989_DzhilkibaevR_YadFiz49}
\bibinfo{author}{\bibfnamefont{R.~M.} \bibnamefont{Dzhilkibaev}} \bibnamefont{and} \bibinfo{author}{\bibfnamefont{V.~M.} \bibnamefont{Lobashev}}, \bibinfo{journal}{Yadernaya Fizika} \textbf{\bibinfo{volume}{49}}, \bibinfo{pages}{622} (\bibinfo{year}{1989}), \bibinfo{note}{[Soviet Journal of Nuclear Physics {\bf{49}}, 384 (1989)]}.

\bibitem[{\citenamefont{Hill et~al.}(2024)\citenamefont{Hill, Plestid, and Zupan}}]{2024_HillR_PRD109}
\bibinfo{author}{\bibfnamefont{R.~J.} \bibnamefont{Hill}}, \bibinfo{author}{\bibfnamefont{R.}~\bibnamefont{Plestid}}, \bibnamefont{and} \bibinfo{author}{\bibfnamefont{J.}~\bibnamefont{Zupan}}, \bibinfo{journal}{Physical Review D} \textbf{\bibinfo{volume}{109}}, \bibinfo{pages}{035025} (\bibinfo{year}{2024}).

\bibitem[{\citenamefont{Tiomno and Wheeler}(1949)}]{1949_TiomnoJ_RMP21}
\bibinfo{author}{\bibfnamefont{J.}~\bibnamefont{Tiomno}} \bibnamefont{and} \bibinfo{author}{\bibfnamefont{J.~A.} \bibnamefont{Wheeler}}, \bibinfo{journal}{Reviews of Modern Physics} \textbf{\bibinfo{volume}{21}}, \bibinfo{pages}{144} (\bibinfo{year}{1949}).

\bibitem[{\citenamefont{Porter and Primakoff}(1951)}]{1951_PorterC_PR83}
\bibinfo{author}{\bibfnamefont{C.~E.} \bibnamefont{Porter}} \bibnamefont{and} \bibinfo{author}{\bibfnamefont{H.}~\bibnamefont{Primakoff}}, \bibinfo{journal}{Physical Review} \textbf{\bibinfo{volume}{83}}, \bibinfo{pages}{849} (\bibinfo{year}{1951}).

\bibitem[{\citenamefont{Tenaglia}(1959)}]{1959_TenagliaL_NouvCim13}
\bibinfo{author}{\bibfnamefont{L.}~\bibnamefont{Tenaglia}}, \bibinfo{journal}{Il Nuovo Cimento} \textbf{\bibinfo{volume}{13}}, \bibinfo{pages}{284} (\bibinfo{year}{1959}).

\bibitem[{\citenamefont{Gilinsky and Mathews}(1960)}]{1960_GilinskyV_PR120}
\bibinfo{author}{\bibfnamefont{V.}~\bibnamefont{Gilinsky}} \bibnamefont{and} \bibinfo{author}{\bibfnamefont{J.}~\bibnamefont{Mathews}}, \bibinfo{journal}{Physical Review} \textbf{\bibinfo{volume}{120}}, \bibinfo{pages}{1450} (\bibinfo{year}{1960}).

\bibitem[{\citenamefont{Überall}(1960)}]{1960_UeberallH_PR119}
\bibinfo{author}{\bibfnamefont{H.}~\bibnamefont{Überall}}, \bibinfo{journal}{Physical Review} \textbf{\bibinfo{volume}{119}}, \bibinfo{pages}{365} (\bibinfo{year}{1960}).

\bibitem[{\citenamefont{Huff}(1961)}]{1961_HuffR_AnnPhys16}
\bibinfo{author}{\bibfnamefont{R.~W.} \bibnamefont{Huff}}, \bibinfo{journal}{Annals of Physics} \textbf{\bibinfo{volume}{16}}, \bibinfo{pages}{288} (\bibinfo{year}{1961}).

\bibitem[{\citenamefont{Johnson et~al.}(1961)\citenamefont{Johnson, O'Connell, and Mullin}}]{1961_JohnsonW_PR124}
\bibinfo{author}{\bibfnamefont{W.~R.} \bibnamefont{Johnson}}, \bibinfo{author}{\bibfnamefont{R.~F.} \bibnamefont{O'Connell}}, \bibnamefont{and} \bibinfo{author}{\bibfnamefont{C.~J.} \bibnamefont{Mullin}}, \bibinfo{journal}{Physical Review} \textbf{\bibinfo{volume}{124}}, \bibinfo{pages}{904} (\bibinfo{year}{1961}).

\bibitem[{\citenamefont{von Baeyer and Leiter}(1979)}]{1979_vonBaeyerH_PRA19}
\bibinfo{author}{\bibfnamefont{H.~C.} \bibnamefont{von Baeyer}} \bibnamefont{and} \bibinfo{author}{\bibfnamefont{D.}~\bibnamefont{Leiter}}, \bibinfo{journal}{Physical Review A} \textbf{\bibinfo{volume}{19}}, \bibinfo{pages}{1371} (\bibinfo{year}{1979}).

\bibitem[{\citenamefont{Hänggi et~al.}(1974)\citenamefont{Hänggi, Viollier, Raff, and Alder}}]{1974_HaenggiP_PhysLettB51}
\bibinfo{author}{\bibfnamefont{P.}~\bibnamefont{Hänggi}}, \bibinfo{author}{\bibfnamefont{R.~D.} \bibnamefont{Viollier}}, \bibinfo{author}{\bibfnamefont{U.}~\bibnamefont{Raff}}, \bibnamefont{and} \bibinfo{author}{\bibfnamefont{K.}~\bibnamefont{Alder}}, \bibinfo{journal}{Physics Letters B} \textbf{\bibinfo{volume}{51}}, \bibinfo{pages}{119} (\bibinfo{year}{1974}).

\bibitem[{\citenamefont{Herzog and Alder}(1980)}]{1980_HerzogF_HevlPhysActa53}
\bibinfo{author}{\bibfnamefont{F.}~\bibnamefont{Herzog}} \bibnamefont{and} \bibinfo{author}{\bibfnamefont{K.}~\bibnamefont{Alder}}, \bibinfo{journal}{Helvetica Physica Acta} \textbf{\bibinfo{volume}{53}} (\bibinfo{year}{1980}).

\bibitem[{\citenamefont{Chatterjee and Roy}(1980)}]{1980_ChatterjeeL_AnnDerPhys492}
\bibinfo{author}{\bibfnamefont{L.}~\bibnamefont{Chatterjee}} \bibnamefont{and} \bibinfo{author}{\bibfnamefont{T.}~\bibnamefont{Roy}}, \bibinfo{journal}{Annalen der Physik} \textbf{\bibinfo{volume}{492}}, \bibinfo{pages}{321} (\bibinfo{year}{1980}).

\bibitem[{\citenamefont{Shanker}(1982)}]{1982_ShankerO_PRD25}
\bibinfo{author}{\bibfnamefont{O.}~\bibnamefont{Shanker}}, \bibinfo{journal}{Physical Review D} \textbf{\bibinfo{volume}{25}}, \bibinfo{pages}{1847} (\bibinfo{year}{1982}).

\bibitem[{\citenamefont{Watanabe et~al.}(1987)\citenamefont{Watanabe, Fukui, Ohtsubo, and Morita}}]{1987_WatanabeR_ProgTheorPhys78}
\bibinfo{author}{\bibfnamefont{R.}~\bibnamefont{Watanabe}}, \bibinfo{author}{\bibfnamefont{M.}~\bibnamefont{Fukui}}, \bibinfo{author}{\bibfnamefont{H.}~\bibnamefont{Ohtsubo}}, \bibnamefont{and} \bibinfo{author}{\bibfnamefont{M.}~\bibnamefont{Morita}}, \bibinfo{journal}{Progress of Theoretical Physics} \textbf{\bibinfo{volume}{78}}, \bibinfo{pages}{114} (\bibinfo{year}{1987}).

\bibitem[{\citenamefont{Watanabe et~al.}(1993)\citenamefont{Watanabe, Muto, Oda, Niwa, Ohtsubo, Morita, and Morita}}]{1993_WatanabeR_ADNDT54}
\bibinfo{author}{\bibfnamefont{R.}~\bibnamefont{Watanabe}}, \bibinfo{author}{\bibfnamefont{K.}~\bibnamefont{Muto}}, \bibinfo{author}{\bibfnamefont{T.}~\bibnamefont{Oda}}, \bibinfo{author}{\bibfnamefont{T.}~\bibnamefont{Niwa}}, \bibinfo{author}{\bibfnamefont{H.}~\bibnamefont{Ohtsubo}}, \bibinfo{author}{\bibfnamefont{R.}~\bibnamefont{Morita}}, \bibnamefont{and} \bibinfo{author}{\bibfnamefont{M.}~\bibnamefont{Morita}}, \bibinfo{journal}{Atomic Data and Nuclear Data Tables} \textbf{\bibinfo{volume}{54}}, \bibinfo{pages}{165} (\bibinfo{year}{1993}).

\bibitem[{\citenamefont{Czarnecki et~al.}(2011)\citenamefont{Czarnecki, Garcia I~Tormo, and Marciano}}]{2011_CzarneckiA_PRD84}
\bibinfo{author}{\bibfnamefont{A.}~\bibnamefont{Czarnecki}}, \bibinfo{author}{\bibfnamefont{X.}~\bibnamefont{Garcia I~Tormo}}, \bibnamefont{and} \bibinfo{author}{\bibfnamefont{W.~J.} \bibnamefont{Marciano}}, \bibinfo{journal}{Physical Review D} \textbf{\bibinfo{volume}{84}}, \bibinfo{pages}{013006} (\bibinfo{year}{2011}).

\bibitem[{\citenamefont{Szafron and Czarnecki}(2016{\natexlab{a}})}]{2016_SzafronR_PhysLettB753}
\bibinfo{author}{\bibfnamefont{R.}~\bibnamefont{Szafron}} \bibnamefont{and} \bibinfo{author}{\bibfnamefont{A.}~\bibnamefont{Czarnecki}}, \bibinfo{journal}{Physics Letters B} \textbf{\bibinfo{volume}{753}}, \bibinfo{pages}{61} (\bibinfo{year}{2016}{\natexlab{a}}).

\bibitem[{\citenamefont{Szafron and Czarnecki}(2016{\natexlab{b}})}]{2016_SzafronR_PRD94}
\bibinfo{author}{\bibfnamefont{R.}~\bibnamefont{Szafron}} \bibnamefont{and} \bibinfo{author}{\bibfnamefont{A.}~\bibnamefont{Czarnecki}}, \bibinfo{journal}{Physical Review D} \textbf{\bibinfo{volume}{94}}, \bibinfo{pages}{051301} (\bibinfo{year}{2016}{\natexlab{b}}).

\bibitem[{\citenamefont{Heeck et~al.}(2022)\citenamefont{Heeck, Szafron, and Uesaka}}]{2022_HeeckJ_PRD105}
\bibinfo{author}{\bibfnamefont{J.}~\bibnamefont{Heeck}}, \bibinfo{author}{\bibfnamefont{R.}~\bibnamefont{Szafron}}, \bibnamefont{and} \bibinfo{author}{\bibfnamefont{Y.}~\bibnamefont{Uesaka}}, \bibinfo{journal}{Physical Review D} \textbf{\bibinfo{volume}{105}}, \bibinfo{pages}{053006} (\bibinfo{year}{2022}).

\bibitem[{\citenamefont{Salvat et~al.}(1995)\citenamefont{Salvat, Fernández-Varea, and Williamson}}]{1995_SalvatF_CPC90}
\bibinfo{author}{\bibfnamefont{F.}~\bibnamefont{Salvat}}, \bibinfo{author}{\bibfnamefont{J.~M.} \bibnamefont{Fernández-Varea}}, \bibnamefont{and} \bibinfo{author}{\bibfnamefont{W.}~\bibnamefont{Williamson}}, \bibinfo{journal}{Computer Physics Communications} \textbf{\bibinfo{volume}{90}}, \bibinfo{pages}{151} (\bibinfo{year}{1995}).

\bibitem[{\citenamefont{Salvat and Fernández-Varea}(2019)}]{2019_SalvatF_CPC240}
\bibinfo{author}{\bibfnamefont{F.}~\bibnamefont{Salvat}} \bibnamefont{and} \bibinfo{author}{\bibfnamefont{J.~M.} \bibnamefont{Fernández-Varea}}, \bibinfo{journal}{Computer Physics Communications} \textbf{\bibinfo{volume}{240}}, \bibinfo{pages}{165} (\bibinfo{year}{2019}).

\bibitem[{\citenamefont{I.Angeli and K.P.Marinova}(2013)}]{Ang13}
\bibinfo{author}{\bibnamefont{I.Angeli}} \bibnamefont{and} \bibinfo{author}{\bibnamefont{K.P.Marinova}}, \bibinfo{journal}{Atomic Data and Nuclear Data Tables} \textbf{\bibinfo{volume}{99}}, \bibinfo{pages}{69} (\bibinfo{year}{2013}).

\bibitem[{\citenamefont{Kozhedub et~al.}(2008)\citenamefont{Kozhedub, Andreev, Shabaev, Tupitsyn, Brandau, Kozhuharov, Plunien, and Stöhlker}}]{2008_KozhedubY_PRA77}
\bibinfo{author}{\bibfnamefont{Y.~S.} \bibnamefont{Kozhedub}}, \bibinfo{author}{\bibfnamefont{O.~V.} \bibnamefont{Andreev}}, \bibinfo{author}{\bibfnamefont{V.~M.} \bibnamefont{Shabaev}}, \bibinfo{author}{\bibfnamefont{I.~I.} \bibnamefont{Tupitsyn}}, \bibinfo{author}{\bibfnamefont{C.}~\bibnamefont{Brandau}}, \bibinfo{author}{\bibfnamefont{C.}~\bibnamefont{Kozhuharov}}, \bibinfo{author}{\bibfnamefont{G.}~\bibnamefont{Plunien}}, \bibnamefont{and} \bibinfo{author}{\bibfnamefont{T.}~\bibnamefont{Stöhlker}}, \bibinfo{journal}{Physical Review A} \textbf{\bibinfo{volume}{77}}, \bibinfo{pages}{032501} (\bibinfo{year}{2008}).

\bibitem[{\citenamefont{Pritychenko et~al.}(2016)\citenamefont{Pritychenko, Birch, Singh, and Horoi}}]{2016_PritychenkoB_ADNDT107}
\bibinfo{author}{\bibfnamefont{B.}~\bibnamefont{Pritychenko}}, \bibinfo{author}{\bibfnamefont{M.}~\bibnamefont{Birch}}, \bibinfo{author}{\bibfnamefont{B.}~\bibnamefont{Singh}}, \bibnamefont{and} \bibinfo{author}{\bibfnamefont{M.}~\bibnamefont{Horoi}}, \bibinfo{journal}{Atomic Data and Nuclear Data Tables} \textbf{\bibinfo{volume}{107}}, \bibinfo{pages}{1} (\bibinfo{year}{2016}).

\bibitem[{\citenamefont{Shabaev et~al.}(2013)\citenamefont{Shabaev, Tupitsyn, and Yerokhin}}]{2013_ShabaevV_PhysRevA}
\bibinfo{author}{\bibfnamefont{V.~M.} \bibnamefont{Shabaev}}, \bibinfo{author}{\bibfnamefont{I.~I.} \bibnamefont{Tupitsyn}}, \bibnamefont{and} \bibinfo{author}{\bibfnamefont{V.~A.} \bibnamefont{Yerokhin}}, \bibinfo{journal}{Physical Review A} \textbf{\bibinfo{volume}{88}}, \bibinfo{pages}{012513} (\bibinfo{year}{2013}).

\bibitem[{\citenamefont{Shabaev et~al.}(2015)\citenamefont{Shabaev, Tupitsyn, and Yerokhin}}]{2015_ShabaevV_CompPhysComm}
\bibinfo{author}{\bibfnamefont{V.~M.} \bibnamefont{Shabaev}}, \bibinfo{author}{\bibfnamefont{I.~I.} \bibnamefont{Tupitsyn}}, \bibnamefont{and} \bibinfo{author}{\bibfnamefont{V.~A.} \bibnamefont{Yerokhin}}, \bibinfo{journal}{Computer Physics Communications} \textbf{\bibinfo{volume}{189}}, \bibinfo{pages}{175} (\bibinfo{year}{2015}).

\bibitem[{\citenamefont{Shabaev et~al.}(2018)\citenamefont{Shabaev, Tupitsyn, and Yerokhin}}]{2018_ShabaevV_CompPhysComm}
\bibinfo{author}{\bibfnamefont{V.~M.} \bibnamefont{Shabaev}}, \bibinfo{author}{\bibfnamefont{I.~I.} \bibnamefont{Tupitsyn}}, \bibnamefont{and} \bibinfo{author}{\bibfnamefont{V.~A.} \bibnamefont{Yerokhin}}, \bibinfo{journal}{Computer Physics Communications} \textbf{\bibinfo{volume}{223}}, \bibinfo{pages}{69} (\bibinfo{year}{2018}).

\bibitem[{\citenamefont{Fullerton and Rinker}(1976)}]{1976_FullertonL_PRA13}
\bibinfo{author}{\bibfnamefont{L.~W.} \bibnamefont{Fullerton}} \bibnamefont{and} \bibinfo{author}{\bibfnamefont{G.~A.} \bibnamefont{Rinker}}, \bibinfo{journal}{Physical Review A} \textbf{\bibinfo{volume}{13}}, \bibinfo{pages}{1283} (\bibinfo{year}{1976}).

\bibitem[{\citenamefont{Fainshtein et~al.}(1991)\citenamefont{Fainshtein, Manakov, and Nekipelov}}]{1991_FainshteinA_JPB24}
\bibinfo{author}{\bibfnamefont{A.~G.} \bibnamefont{Fainshtein}}, \bibinfo{author}{\bibfnamefont{N.~L.} \bibnamefont{Manakov}}, \bibnamefont{and} \bibinfo{author}{\bibfnamefont{A.~A.} \bibnamefont{Nekipelov}}, \bibinfo{journal}{Journal of Physics B: Atomic, Molecular and Optical Physics} \textbf{\bibinfo{volume}{24}}, \bibinfo{pages}{559} (\bibinfo{year}{1991}).

\bibitem[{\citenamefont{Yerokhin and Oreshkina}(2023)}]{2023_YerokhinV_PRA108}
\bibinfo{author}{\bibfnamefont{V.~A.} \bibnamefont{Yerokhin}} \bibnamefont{and} \bibinfo{author}{\bibfnamefont{N.~S.} \bibnamefont{Oreshkina}}, \bibinfo{journal}{Physical Review A} \textbf{\bibinfo{volume}{108}}, \bibinfo{pages}{052824} (\bibinfo{year}{2023}).

\bibitem[{\citenamefont{Wang et~al.}(2021)\citenamefont{Wang, Huang, Kondev, Audi, and Naimi}}]{2021_WangM_ChinPhysC45}
\bibinfo{author}{\bibfnamefont{M.}~\bibnamefont{Wang}}, \bibinfo{author}{\bibfnamefont{W.}~\bibnamefont{Huang}}, \bibinfo{author}{\bibfnamefont{F.}~\bibnamefont{Kondev}}, \bibinfo{author}{\bibfnamefont{G.}~\bibnamefont{Audi}}, \bibnamefont{and} \bibinfo{author}{\bibfnamefont{S.}~\bibnamefont{Naimi}}, \bibinfo{journal}{Chinese Physics C} \textbf{\bibinfo{volume}{45}}, \bibinfo{pages}{030003} (\bibinfo{year}{2021}).

\bibitem[{\citenamefont{Sapirstein and Cheng}(2002)}]{2002_SapirsteinJ_PRA66}
\bibinfo{author}{\bibfnamefont{J.}~\bibnamefont{Sapirstein}} \bibnamefont{and} \bibinfo{author}{\bibfnamefont{K.~T.} \bibnamefont{Cheng}}, \bibinfo{journal}{Physical Review A} \textbf{\bibinfo{volume}{66}}, \bibinfo{pages}{042501} (\bibinfo{year}{2002}).

\bibitem[{\citenamefont{Borie and Rinker}(1982)}]{1982_BorieE_RMP54}
\bibinfo{author}{\bibfnamefont{E.}~\bibnamefont{Borie}} \bibnamefont{and} \bibinfo{author}{\bibfnamefont{G.~A.} \bibnamefont{Rinker}}, \bibinfo{journal}{Reviews of Modern Physics} \textbf{\bibinfo{volume}{54}}, \bibinfo{pages}{67} (\bibinfo{year}{1982}).

\bibitem[{\citenamefont{Varshalovich et~al.}(1988)\citenamefont{Varshalovich, Moskalev, and Khersonskii}}]{1988_VarshalovichD_QTAM}
\bibinfo{author}{\bibfnamefont{D.~A.} \bibnamefont{Varshalovich}}, \bibinfo{author}{\bibfnamefont{A.~N.} \bibnamefont{Moskalev}}, \bibnamefont{and} \bibinfo{author}{\bibfnamefont{V.~K.} \bibnamefont{Khersonskii}}, \emph{\bibinfo{title}{Quantum Theory of Angular Momentum}} (\bibinfo{publisher}{World Scientific}, \bibinfo{address}{Singapore}, \bibinfo{year}{1988}).

\bibitem[{\citenamefont{Grant}(1970)}]{1970_GrantI_AP19}
\bibinfo{author}{\bibfnamefont{I.}~\bibnamefont{Grant}}, \bibinfo{journal}{Advances in Physics} \textbf{\bibinfo{volume}{19}}, \bibinfo{pages}{747} (\bibinfo{year}{1970}).

\end{thebibliography}

\newpage
\appendix

\onecolumngrid
\section{Multipole expansion of the operator~$V^{\mathrm{F}}\left(1,2\right)$}\label{app:a}

The multipole expansion of the two-particle interaction operator $V^{\mathrm{F}}$ defined in Eq.~\eqref{eq:fermi_interaction_operator} can be derived from the multipole expansion of the delta function $\delta\left( \vec{r}_1 - \vec{r}_2 \right)$:
\begin{equation}
\delta\left(\vec{r}_{1}-\vec{r}_{2}\right)=\sum_{lm}u_{l}\left(r_{1},r_{2}\right)C_{l}^{m}\left(1\right)C_{lm}\left(2\right),\qquad u_{l} \left(r_1,r_2\right)=\frac{\Pi_{l}^{2}}{4\pi}\frac{\delta\left(r_{1}-r_{2}\right)}{r_1 r_2},
\end{equation}
where
\begin{equation}\label{eq:pi}
    \Pi_{l_{1}\dots l_{n}}=\sqrt{2l_{1}+1}\dots\sqrt{2l_{n}+1},
\end{equation}
\begin{equation}
    C_{lm}\left(j\right)=\frac{\sqrt{4\pi}}{\Pi_{l}}Y_{lm}\left(j\right)
\end{equation}
is the spherical tensor of the rank $l$, $Y_{lm}$ is the spherical harmonic, $r = \left| \vec{r} \right|$.
Here and hereafter, the sum over $l$ runs from zero to infinity and the sum over $m$ runs in the range $-l\leq m \leq l$, if not specified otherwise.
\par
Instead of the Dirac $\gamma$ matrices, we employ the~$\alpha$ and $\Sigma$ matrices, which are~$\alpha^{\rho} = \gamma^0 \gamma^{\rho}$ and $\Sigma^{\rho} = \alpha^{\rho}\gamma^5$.
We consider separately the contributions of the time- and space-like components of the corresponding operators:~$\left(\alpha^0,\vec{\alpha}\right)$ and~$\left(\Sigma^0,\vec{\Sigma} \right)$ and refer to them as the scalar and vector contributions.
We treat the vector-like operators as the spherical tensors of the rank 1.
Let us separate the scalar,
\begin{equation}\label{eq:fermi_interaction_operator_s}
    V^{\mathrm{s}}_l\left(1,2\right) = u_{l}\left(r_{1},r_{2}\right) \sum_m  \left[\alpha^{0}-\Sigma^{0}\right]\left(1\right)\left[\alpha_{0}-\Sigma_{0}\right]\left(2\right) C_{l}^{m}\left(1\right)C_{lm}\left(2\right),
\end{equation}
and the vector,
\begin{equation}\label{eq:fermi_interaction_operator_v}
V^{\mathrm{v}}_l\left(1,2\right) = u_{l}\left(r_{1},r_{2}\right) \sum_{mi} \left[\alpha_{1}^{i}-\Sigma^{i}_{1}\right]\left(1\right)\left[\alpha_{1i}-\Sigma_{1i}\right]\left(2\right) C_{l}^{m}\left(1\right)C_{lm}\left(2\right) ,
\end{equation}
parts of the Lorentz scalar in Eq.~\eqref{eq:fermi_interaction_operator}.
In Eq.~\eqref{eq:fermi_interaction_operator_v}, the index~$i$ runs over the set $\left(1,2,3\right)$ and subscripts ``1'' for $\alpha$ and $\Sigma$ refer to the spherical-tensor rank.
Introducing some auxiliary operators,
\begin{equation}\label{eq:o1-o4}
\begin{aligned}O_{l}^{1}\left(1,2\right)=\sum_{m}C_{l}^{m}\left(1\right)C_{lm}\left(2\right), & \qquad O_{l}^{2}\left(1,2\right)=\sum_{m}\left[\Sigma_{0}C_{l}^{m}\right]\left(1\right)C_{lm}\left(2\right),\\
O_{l}^{3}\left(1,2\right)=\sum_{m}C_{l}^{m}\left(1\right)\left[\Sigma_{0}C_{lm}\right]\left(2\right), & \qquad O_{l}^{4}\left(1,2\right)=\sum_{m}\left[\Sigma_{0}C_{l}^{m}\right]\left(1\right)\left[\Sigma_{0}C_{lm}\right]\left(2\right),
\end{aligned}
\end{equation}
the scalar part of the interaction operator, Eq.~\eqref{eq:fermi_interaction_operator_s}, can be written as
\begin{equation}\label{eq:fermi_interaction_operator_s_o}
V_l^{\mathrm{s}}\left(1,2\right) = u_{l}\left(r_{1},r_{2}\right)\left[O_{l}^{1}\left(1,2\right)-O_{l}^{2}\left(1,2\right)-O_{l}^{3}\left(1,2\right)+O_{l}^{4}\left(1,2\right)\right].
\end{equation}
Changing the coupling scheme of scalar products in Eq.~\eqref{eq:fermi_interaction_operator_v}~\cite{1988_VarshalovichD_QTAM} and introducing another set of auxiliary operators
\begin{equation}
\begin{aligned}O_{l}^{5}\left(1,2\right)=\sum_{L=l-1}^{l+1}\left(-1\right)^{L-l+1}\left(\left[\alpha_{1}\left(1\right)\otimes C_{l}\left(1\right)\right]_{L}\cdot\left[\alpha_{1}\left(2\right)\otimes C_{l}\left(2\right)\right]_{L}\right) & ,\\
O_{l}^{6}\left(1,2\right)=\sum_{L=l-1}^{l+1}\left(-1\right)^{L-l+1}\left(\left[\Sigma_{1}\left(1\right)\otimes C_{l}\left(1\right)\right]_{L}\cdot\left[\alpha_{1}\left(2\right)\otimes C_{l}\left(2\right)\right]_{L}\right) & ,\\
O_{l}^{7}\left(1,2\right)=\sum_{L=l-1}^{l+1}\left(-1\right)^{L-l+1}\left(\left[\alpha_{1}\left(1\right)\otimes C_{l}\left(1\right)\right]_{L}\cdot\left[\Sigma_{1}\left(2\right)\otimes C_{l}\left(2\right)\right]_{L}\right) & ,\\
O_{l}^{8}\left(1,2\right)=\sum_{L=l-1}^{l+1}\left(-1\right)^{L-l+1}\left(\left[\Sigma_{1}\left(1\right)\otimes C_{l}\left(1\right)\right]_{L}\cdot\left[\Sigma_{1}\left(2\right)\otimes C_{l}\left(2\right)\right]_{L}\right) & ,
\end{aligned}
\end{equation}
where~$\left[A \otimes B\right]_{LM}$ denotes the irreducible tensor product of operators, allows us to rewrite the vector part, Eq.~\eqref{eq:fermi_interaction_operator_v}, as
\begin{equation}\label{eq:fermi_interaction_operator_v_o}
V_l^{\mathrm{v}}\left(1,2\right) =  u_{l}\left(r_{1},r_{2}\right)\left[O_{l}^{5}\left(1,2\right)-O_{l}^{6}\left(1,2\right)-O_{l}^{7}\left(1,2\right)+O_{l}^{8}\left(1,2\right)\right].
\end{equation}

\onecolumngrid
\section{Relativistic matrix elements of the operator~$V^{\mathrm{F}}\left(1,2\right)$ in the central-field approximation}\label{app:b}

The relativistic four-component wave function in the central-field approximation can be represented in the form~\cite{1970_GrantI_AP19}
\begin{equation}\label{eq:wf_centralf_field}
\psi_{n\varkappa\mu}\left(\vec{r},\sigma\right)=\left(\begin{array}{c}
\psi_{n\varkappa\mu}^{\beta=+1}\left(\vec{r},\sigma\right)\\
\psi_{n\varkappa\mu}^{\beta=-1}\left(\vec{r},\sigma\right)
\end{array}\right),
\end{equation}
where~$n$ is the principal quantum number in the bound-state case (it should be replaced with the energy~$E$ in the unbound-state case), and the index~$\beta=\left(+1,-1\right)$ enumerates the conventional large,~$\beta=+1$, and small, $\beta=-1$, components of the wave function.
In turn, the large and small components read as follows:
\begin{equation}\label{eq:wf_centralf_field_beta}
    \psi_{n\varkappa\mu}^{\beta}\left(\vec{r},\sigma\right)=i^{\frac{\left(1-\beta\right)}{2}}\frac{F_{n\varkappa}^{\beta}\left(r\right)}{r}\chi_{\varkappa\mu}^{\beta}\left(\Omega,\sigma\right),
\end{equation}
where
\begin{equation}\label{eq:spinor_centralf_field_beta}
\chi_{\varkappa\mu}^{\beta}\left(\Omega,\sigma\right)=\begin{cases}
\chi_{+\varkappa\mu}\left(\Omega,\sigma\right), & \beta=+1,\\
\chi_{-\varkappa\mu}\left(\Omega,\sigma\right), & \beta=-1,
\end{cases}
\end{equation}
and $\chi_{\varkappa\mu}\left(\Omega,\sigma\right)$ is the Pauli spherical spinor.
\par
Using Eqs.~\eqref{eq:o1-o4} and \eqref{eq:fermi_interaction_operator_s_o} and our definition of the wave function,~\eqref{eq:wf_centralf_field} --~\eqref{eq:spinor_centralf_field_beta}, for the two-particle matrix elements of the operator $V^{\mathrm{s}}_l\left(1,2\right)$ we obtain
\begin{equation}\label{eq:vs_l_abcd}
\braket{cd|V_{l}^{\mathrm{\mathrm{s}}}\left(1,2\right)|ab} = \frac{\Pi_l^2}{4\pi} \sum_{m}  g^{lm}\left(j_{c}\mu_{c};j_{a}\mu_{a}\right)g^{lm}\left(j_{b}\mu_{b};j_{d}\mu_{d}\right)R_{l}^{\mathrm{s}}\left(cd,ab\right),
\end{equation}
where $g^{lm}\left(j_{1}\mu_{1};j_{2}\mu_{2}\right)$ are the relativistic Gaunt coefficients,
\begin{equation}
    g^{lm}\left(j_{1}\mu_{1};j_{2}\mu_{2}\right) = \braket{\varkappa_1\mu_1|C_{lm}|\varkappa_2\mu_2},
\end{equation}
stemmed from the spin-angular integrations, and the radial part is
\begin{equation}\label{eq:2p_rads_cdab}
\begin{aligned}
R_{l}^{\mathrm{s}}\left(cd,ab\right) = & \sum_{\beta_{1}\beta_{2}}\bigg\{\left[E\left(l,l_{c},l_{a}\right)E\left(l,l_{b},l_{d}\right)R_{1}^{\beta_{1}\beta_{2}}\left(cd,ab\right)-\beta_{1}\beta_{2}O\left(l,l_{c},l_{a}\right)O\left(l,l_{b},l_{d}\right)R_{4}^{\beta_{1}\beta_{2}}\left(cd,ab\right)\right]\\
 & \quad\,\,\,-i\left[\beta_{1}O\left(l,l_{c},l_{a}\right)E\left(l,l_{b},l_{d}\right)R_{2}^{\beta_{1}\beta_{2}}\left(cd,ab\right)+\beta_{2}E\left(l,l_{c},l_{a}\right)O\left(l,l_{b},l_{d}\right)R_{3}^{\beta_{1}\beta_{2}}\left(cd,ab\right)\right]\bigg\},
\end{aligned}
\end{equation}
where the functions
\begin{equation}
E\left(q,w,e\right)=\begin{cases}
1, & q+w+e\mathrm{\,\,is\,\,\,even},\\
0, & q+w+e\mathrm{\,\,is\,\,\,odd},
\end{cases}\qquad O\left(q,w,e\right)=\begin{cases}
0, & q+w+e \mathrm{\,\,is\,\,\,even},\\
1, & q+w+e \mathrm{\,\,is\,\,\,odd},
\end{cases}
\end{equation}
are introduced.
The two-particle radial integrals $R_{1-4}^{\beta_{1}\beta_{2}}\left(cd,ab\right)$  in Eq.~\eqref{eq:vs_l_abcd} are given by
\begin{equation}\label{eq:two_particle_radints}
\begin{aligned}
R_{1}^{\beta_{1}\beta_{2}}\left(cd,ab\right)= & \int dr\, \left[ F_{n_{c}\varkappa_{c}}^{\beta_{1}}\left(r\right)F_{n_{d}\varkappa_{d}}^{\beta_{2}}\left(r\right) \right]^*\frac{1}{r^{2}}F_{n_{a}\varkappa_{a}}^{\beta_{1}}\left(r\right)F_{n_{b}\varkappa_{b}}^{\beta_{2}}\left(r\right),\\
R_{2}^{\beta_{1}\beta_{2}}\left(cd,ab\right)= & \int dr\, \left[ F_{n_{c}\varkappa_{c}}^{\beta_{1}}\left(r\right)F_{n_{d}\varkappa_{d}}^{\beta_{2}}\left(r\right) \right]^*\frac{1}{r^{2}}F_{n_{a}\varkappa_{a}}^{-\beta_{1}}\left(r\right)F_{n_{b}\varkappa_{b}}^{\beta_{2}}\left(r\right),\\
R_{3}^{\beta_{1}\beta_{2}}\left(cd,ab\right)= & \int dr\, \left[ F_{n_{c}\varkappa_{c}}^{\beta_{1}}\left(r\right)F_{n_{d}\varkappa_{d}}^{\beta_{2}}\left(r\right) \right]^*\frac{1}{r^{2}}F_{n_{a}\varkappa_{a}}^{\beta_{1}}\left(r\right)F_{n_{b}\varkappa_{b}}^{-\beta_{2}}\left(r\right),\\
R_{4}^{\beta_{1}\beta_{2}}\left(cd,ab\right)= & \int dr\, \left[ F_{n_{c}\varkappa_{c}}^{\beta_{1}}\left(r\right)F_{n_{d}\varkappa_{d}}^{\beta_{2}}\left(r\right) \right]^*\frac{1}{r^{2}}F_{n_{a}\varkappa_{a}}^{-\beta_{1}}\left(r\right)F_{n_{b}\varkappa_{b}}^{-\beta_{2}}\left(r\right).
\end{aligned}
\end{equation}
\par
The matrix elements of the operator $V^{\mathrm{v}}_l\left(1,2\right)$ can be evaluated using the spin-angular matrix elements of the operator $[ \sigma_1\otimes C_l]_{LM}$ with the functions $\chi_{\varkappa\mu}^{\beta}\left(\Omega,\sigma\right)$~\cite{1970_GrantI_AP19}:
\begin{equation}
\braket{\beta\varkappa_{b}\mu_{b}|\left[\sigma_{1}\otimes C_{l}\right]_{LM}|\beta\varkappa_{a}\mu_{a}} = \beta E\left(l,l_{b},l_{a}\right) g^{LM}\left(j_{b}\mu_{b};j_{a}\mu_{a}\right)S_{l}^{\beta}\left(\varkappa_{b};\varkappa_{a}|L\right),
\end{equation}
\begin{equation}
\braket{\beta\varkappa_{b}\mu_{b}|\left[\sigma_{1}\otimes C_{l}\right]_{LM}|-\beta\varkappa_{a}\mu_{a}}=\beta O\left(l,l_{b},l_{a}\right) g^{LM}\left(j_{b}\mu_{b};j_{a}\mu_{a}\right)U_{l}^{\beta}\left(\varkappa_{b};\varkappa_{a}|L\right).
\end{equation}
These expressions may serve as a definition of the function $U_{l}^{\beta}\left(\varkappa_{b};\varkappa_{a}|L\right)$ and $S_{l}^{\beta}\left(\varkappa_{b};\varkappa_{a}|L\right)$.
One can obtain,
\begin{equation}\label{eq:vv_l_abcd}
\braket{cd|V_{l}^{\mathrm{\mathrm{v}}}\left(1,2\right)|ab} = -\frac{\Pi_l^2}{4\pi} \sum_{LM}  g^{LM}\left(j_{c}\mu_{c};j_{a}\mu_{a}\right)g^{LM}\left(j_{b}\mu_{b};j_{d}\mu_{d}\right)R_{lL}^{\mathrm{v}}\left(cd,ab\right),
\end{equation}
where
\begin{equation}\label{eq:2p_radv_cdab}
\begin{aligned}
R_{lL}^{\mathrm{v}}\left(cd,ab\right)=\sum_{\beta_{1}\beta_{2}} & \bigg\{\Big[\beta_{1}\beta_{2}E\left(l,l_{c},l_{a}\right)E\left(l,l_{b},l_{d}\right)S_{l}^{\beta_{1}}\left(\varkappa_{c},\varkappa_{a}|L\right)S_{l}^{\beta_{2}}\left(\varkappa_{b},\varkappa_{d}|L\right)R_{1}^{\beta_{1}\beta_{2}}\left(cd,ab\right)\\
 & +O\left(l,l_{c},l_{a}\right)O\left(l,l_{b},l_{d}\right)U_{l}^{\beta_{1}}\left(\varkappa_{c},\varkappa_{a}|L\right)U_{l}^{-\beta_{2}}\left(\varkappa_{b},\varkappa_{d}|L\right)R_{4}^{\beta_{1}\beta_{2}}\left(cd,ab\right)\Big]\\
 & +i\Big[\beta_{2}O\left(l,l_{c},l_{a}\right)E\left(l,l_{b},l_{d}\right)U_{l}^{\beta_{1}}\left(\varkappa_{c},\varkappa_{a}|L\right)S_{l}^{\beta_{2}}\left(\varkappa_{b},\varkappa_{d}|L\right)R_{2}^{\beta_{1}\beta_{2}}\left(cd,ab\right)\\
 & -\beta_{1}E\left(l,l_{c},l_{a}\right)O\left(l,l_{b},l_{d}\right)S_{l}^{\beta_{1}}\left(\varkappa_{c},\varkappa_{a}|L\right)U_{l}^{-\beta_{2}}\left(\varkappa_{b},\varkappa_{d}|L\right)R_{3}^{\beta_{1}\beta_{2}}\left(cd,ab\right)\Big]\bigg\}.
\end{aligned}
\end{equation}
\par
Finally, the matrix element of the Fermi's operator can be written as
\begin{equation}\label{eq:vf_cdab}
\braket{cd|V^{\mathrm{F}}\left(1,2\right)|ab}=\frac{G_{\mathrm{F}}}{\sqrt{2}}\frac{1}{4\pi}\sum_{lmLM}\Pi_{l}^{2}g^{LM}\left(j_{c}\mu_{c};j_{a}\mu_{a}\right)g^{LM}\left(j_{b}\mu_{b};j_{d}\mu_{d}\right)\left[\delta_{Ll}\delta_{Mm}R_{l}^{\mathrm{s}}\left(cd,ab\right)+R_{lL}^{\mathrm{v}}\left(cd,ab\right)\right],
\end{equation}
where $\delta_{ab}$ is the Kronecker delta.

\onecolumngrid
\section{Sum over the external total angular-momentum projections in $\overline{|A|^2}$}\label{app:c}

The amplitude summed over the total angular-momentum projections of all particles can be obtained employing the sum rule for the relativistic Gaunt coefficients:
\begin{equation}\label{eq:rgaunt_sumrule}
    \sum_{\mu_{a}\mu_{b}}g^{lm}\left(j_{b}\mu_{b};j_{a}\mu_{a}\right)g^{kq}\left(j_{b}\mu_{b};j_{a}\mu_{a}\right) = \frac{\Pi_{j_{a}}^{2}}{\Pi_{l}^{2}}\left(C_{l0j_{a}\frac{1}{2}}^{j_{b}\frac{1}{2}}\right)^{2}\delta_{lk}\delta_{mq}.
\end{equation}
Applying Eq.~\eqref{eq:rgaunt_sumrule} to Eq.~\eqref{eq:vf_cdab} results in
\begin{equation}
\sum_{\mu_{a}\mu_{b}\mu_{c}\mu_{d}}\left|\braket{cd|V^{\mathrm{F}}\left(1,2\right)|ab}\right|^{2}=\frac{G_{\mathrm{F}}^{2}}{32\pi^2}\Pi_{j_{a}j_{d}}^{2}\sum_{l}\left(\Pi_{l}C_{l0j_{a}\frac{1}{2}}^{j_{c}\frac{1}{2}}C_{l0j_{d}\frac{1}{2}}^{j_{b}\frac{1}{2}}\right)^{2}\left|R_{l}^{\mathrm{s}}\left(cd,ab\right)+\sum_{L}\frac{\Pi_{L}^{2}}{\Pi_{l}^{2}}R_{Ll}^{\mathrm{v}}\left(cd,ab\right)\right|^{2}.
\end{equation}

\onecolumngrid
\section{Separation of the lepton and neutrino degrees of freedom}\label{app:d}
We start from rewriting the expression for the amplitude in the form
\begin{equation}\label{eq:d1}
    A=\frac{G_{\mathrm{F}}}{\sqrt{2}}\int d\vec{r}M^{\rho}\left(e\mu;\vec{r}\right)M_{\rho}\left(\nu_{\mu}\nu_{e};\vec{r}\right),
\end{equation}
where
\begin{equation}
    M^{\rho}\left(ba;\vec{r}\right)=\int dx_{1}\,\delta\left(\vec{r}_{1}-\vec{r}\right)\psi_{b}^{\dagger}\left(x_{1}\right)\gamma^{0}\gamma^{\rho}\left(1-\gamma^{5}\right)\psi_{a}\left(x_{1}\right)
\end{equation}
and $x\equiv\left(\vec{r},\sigma\right)$.
Using the explicit form of relativistic plane waves,
\begin{equation}
    \psi_{E\vec{p}m_{s}}\left(\vec{r},\sigma\right)=\sqrt{\frac{1}{2E}}\frac{e^{i\vec{p}\cdot\vec{r}}}{\left(2\pi\right)^{3/2}}u_{\vec{p}m_{s}}\left(\sigma\right),\qquad\psi_{-E-\vec{p}m_{s}}\left(\vec{r},\sigma\right) = \sqrt{\frac{1}{2E}}\frac{e^{-i\vec{p}\cdot\vec{r}}}{\left(2\pi\right)^{3/2}}v_{\vec{p}m_{s}}\left(\sigma\right),
\end{equation}
where $E = |E| > 0$ and the bispinors are normalized according to $u^{\dagger}_{\vec{p}m_{s}}u_{\vec{p}m_{s}}= 2E$ and $v^{\dagger}_{\vec{p}m_{s}}v_{\vec{p}m_{s}} = 2E$, one can evaluate the radial integral of the neutrino part
\begin{equation}\label{eq:d4}
    M_{\rho}\left(\nu_{\mu}\nu_{e};\vec{r}\right)=\frac{1}{\sqrt{2E_{\nu_{e}}2E_{\nu_{\mu}}}}\frac{1}{\left(2\pi\right)^{3}}\int\,d\vec{k}\,\delta\left(\vec{k}-\vec{p}_{\nu_{\mu}}-\vec{p}_{\nu_{e}}\right)e^{-i\vec{k}\cdot\vec{r}}\sum_{\sigma}\overline{u}_{\vec{p}_{\nu_\mu}m_{s_{\nu_\mu}}}\left(\sigma\right)\gamma^{0}\gamma_{\rho}\left(1-\gamma^{5}\right)v_{\vec{p}_{\nu_e}m_{s_{\nu_e}}}\left(\sigma\right).
\end{equation}
In Eq.~\eqref{eq:d4}, we have introduced a $\delta$ function and related with it integration over some auxiliary three-momentum $\vec{k}$.
Substituting Eq.~\eqref{eq:d4} back to Eq.~\eqref{eq:d1}, the amplitude becomes
\begin{equation}
    A=\frac{G_{\mathrm{F}}}{\sqrt{2}}\frac{1}{\sqrt{2E_{\nu_{e}}2E_{\nu_{\mu}}}}\frac{1}{\left(2\pi\right)^{3}}\int d\vec{k}\,\delta\left(\vec{k}-\vec{p}_{\nu_{\mu}}-\vec{p}_{\nu_{e}}\right)\,J^{\rho}\left(e\mu;\vec{k}\right)\sum_{\sigma}\overline{u}_{\vec{p}_{\mu}m_{s_{\mu}}}\left(\sigma\right)\gamma^{0}\gamma_{\rho}\left(1-\gamma^{5}\right)v_{\vec{p}_{e}m_{s_{e}}}\left(\sigma\right),
\end{equation}
where $J^{\rho}\left(e\mu;\vec{k}\right)$ is the effective lepton current defined in  Eq.~\eqref{eq:effective_current}.
Now we are in position to evaluate the absolute value of the amplitude squared and summed over the spin projections on the direction of $\vec{p}$ of undetected neutrinos
\begin{equation}
    \sum_{m_{s_{e}}m_{s_{\mu}}}\left|A\right|^{2}=\frac{G_{\mathrm{F}}^{2}}{2}\frac{1}{\left(2\pi\right)^{6}}\int\,d\vec{k}\,\delta\left(\vec{k}-\vec{p}_{\nu_{\mu}}-\vec{p}_{\nu_{e}}\right)L_{\alpha\beta}\left(\vec{k}\right)\frac{N^{\alpha\beta}}{4E_{\nu_{\mu}}E_{\nu_{e}}},
\end{equation}
where the lepton tensor is
\begin{equation}
    L_{\alpha\beta}\left(e\mu;\vec{k}\right)=J_{\alpha}\left(e\mu;\vec{k}\right)J_{\beta}^{\dagger}\left(e\mu;\vec{k}\right)
\end{equation}
and the neutrino tensor is
\begin{equation}
    N^{\alpha\beta}=8\left(-g^{\alpha\beta}g_{\rho\tau}p_{\nu_{e}}^{\rho}p_{\nu_{\mu}}^{\tau}+p_{\nu_{e}}^{\alpha}p_{\nu_{\mu}}^{\beta}+p_{\nu_{e}}^{\beta}p_{\nu_{\mu}}^{\alpha}+i\varepsilon^{\alpha\beta\rho\tau}p_{\nu_{e}\rho}p_{\nu_{\mu}\tau}\right).
\end{equation}
\par
To write down the expression for the differential probability, we form the four-dimensional delta function over the intermediate four-momentum~$k$ and obtain
\begin{equation}
    dW\left(\zeta_{e},\zeta_{\mu}\right)=\frac{G_{\mathrm{F}}^{2}}{2}\frac{1}{\left(2\pi\right)^{5}}\int\,dk\,d\vec{p}_{\nu_{e}}\,d\vec{p}_{\nu_{\mu}}\,dE_{e}\,\delta\left(E_{\mu}-E_{e}-k^{0}\right)\delta\left(k-p_{\nu_{\mu}}-p_{\nu_{e}}\right)L_{\alpha\beta}\left(e\mu;\vec{k}\right)\frac{N^{\alpha\beta}}{4E_{\nu_{\mu}}E_{\nu_{e}}}.
\end{equation}
The integration over the three-momenta of neutrinos is performed using the formula
\begin{equation}
    \int d\vec{p}_{\nu_{e}}\,d\vec{p}_{\nu_{\mu}}\,\frac{\delta\left(k-p_{\nu_{\mu}}-p_{\nu_e}\right)N^{\alpha\beta}}{4E_{\nu_{e}}E_{\nu_{\mu}}}=\frac{4\pi}{3}\left(-g^{\alpha\beta}k^{2}+k^{\alpha}k^{\beta}\right).
\end{equation}
Finally, we assemble the expression for the electron spectrum within the one-particle effective approach
\begin{equation}\label{eq:d11}
    \frac{dW\left(E_{e},n_{\mu}\varkappa_{\mu}\right)}{dE_{e}}=\frac{G_{\mathrm{F}}^{2}}{2}\frac{1}{\left(2\pi\right)^{5}}\frac{4\pi}{3}\frac{1}{\Pi_{j_{\mu}}^{2}}\sum_{\mu_{\mu}}\sum_{\varkappa_{e}\mu_{e}}\int\,d\vec{k}\,L_{\alpha\beta}\left(e\mu;\vec{k}\right)\left(-g^{\alpha\beta}k^{2}+k^{\alpha}k^{\beta}\right),
\end{equation}
where $k^{0}=E_{\mu}-E_{e}$.

\onecolumngrid
\section{Multipole expansion and relativistic matrix elements of the effective current in the central-field approximation}\label{app:e}
The multipole expansion of the effective current is based on the following expansion of the exponent
\begin{equation}
    e^{-i\vec{k}\cdot\vec{r}} = \sqrt{4\pi}\sum_{lm}\Pi_{l}i^{-l}j_{l}\left(kr\right)C_{lm}\left(\hat{r}\right)Y_{lm}^{*}\left(\hat{k}\right),
\end{equation}
where $\hat{n} = \vec{n}/|\vec{n}|$.
This leads to
\begin{equation}
    J_{\rho}\left(ba;\vec{k}\right)=\sqrt{4\pi}\sum_{lm}\Pi_{l}i^{-l}Y_{lm}^{*}\left(\hat{k}\right)J_{\rho,lm}\left(ba;k\right),
\end{equation}
where the $lm$-th partial-wave component of the effective current is
\begin{equation}
    J_{\rho,lm}\left(ba;k\right)\equiv\braket{b|\left(\alpha_{\rho}-\Sigma_{\rho}\right)j_{l}\left(kr\right)C_{lm}|a}.
\end{equation}
We emphasize that here $\rho$ is the Lorentz index but $m$ is the cyclic index (it indicates the component of the corresponding spherical tensor).
\par
The matrix elements of the partial-wave components of the effective current can be evaluated using the results of Appendix~\ref{app:b}.
One can obtain for the scalar part, $\rho=0$,
\begin{equation}
J_{0,lm}\left(ba;k\right)=g^{lm}\left(j_{b}\mu_{b};j_{a}\mu_{a}\right)\left[E\left(l_{b},l_{a},l\right)\sum_{\beta}R^{\beta\beta}\left(ba;j_{l}\right)-iO\left(l_{b},l_{a},l\right)\sum_{\beta}\beta R^{\beta-\beta}\left(ba;j_{l}\right)\right]
\end{equation}
and for the vector part, $\rho=t$, $t=1,2,3$,
\begin{equation}
\begin{aligned}J_{t,lm}\left(ba;k\right)= & \sum_{LM}C_{1tlm}^{LM}g^{LM}\left(j_{b}\mu_{b};j_{a}\mu_{a}\right)\times\\
\times & \left[-E\left(l_{b},l_{a},l\right)\sum_{\beta}\beta S_{l}^{\beta}\left(\varkappa_{b},\varkappa_{a}|L\right)R^{\beta\beta}\left(ba;j_{l}\right)+iO\left(l_{b},l_{a},l\right)\sum_{\beta}U_{l}^{\beta}\left(\varkappa_{b},\varkappa_{a}|L\right)R^{\beta-\beta}\left(ba;j_{l}\right)\right],
\end{aligned}
\end{equation}
where the one-particle radial integrals are given by
\begin{equation}
R^{\beta_{1}\beta_{2}}\left(ba;j_{l}\right)=\int_0^{\infty}\left[F_{n_b\varkappa_b}^{\beta_{1}}\left(r\right)\right]^{*} j_{l}\left(kr\right)F_{n_a\varkappa_a}^{\beta_{2}}\left(r\right)\,dr.
\end{equation}

\onecolumngrid
\section{Angular integration the over intermediate momentum in the one-particle effective approach}\label{app:f}
The angular integration in Eq.~\eqref{eq:d11} can be performed for the spherical components of the vector $\vec{k}$ using the Clebsch-Gordan expansion for products of spherical harmonics and the well-known expression for the integrals of products of spherical harmonics over the solid angle.
Executing the above-mentioned steps, we represent the resulting integration over the angular coordinates in the following form
\begin{equation}\label{eq:f1}
\begin{aligned}
\int\,d\vec{k}\,L_{\alpha\beta}\left(ba;\vec{k}\right)\left(-g^{\alpha\beta}k^{2}+k^{\alpha}k^{\beta}\right)=4\pi & \sum_{LM}g^{LM}\left(j_{b}\mu_{b};j_{a}\mu_{a}\right)g^{LM}\left(j_{b}\mu_{b};j_{a}\mu_{a}\right)\\
\times & \int dk\,k^{2}\Big\{ k^{2}R_{L}^{\mathrm{ss}}\left(ba;k\right)-2k^{0}k\mathrm{Re}R_{L}^{\mathrm{sv}}\left(ba;k\right)\\
 & +\left[\left(k^{0}\right)^{2}-k^{2}\right]R_{L}^{\mathrm{vv,1}}\left(ba;k\right)+k^{2}R_{L}^{\mathrm{vv,2}}\left(ba;k\right)\Big\}.
\end{aligned}
\end{equation}
The scalar-scalar (ss), scalar-vector (sv), and vector-vector diagonal, (vv,1) and vector-vector non-diagonal (vv,2) radial integrals are given by
\begin{equation}\label{eq:r_ss_oneparticle}
R_{L}^{\mathrm{ss}}\left(ba;k\right)=\Pi_{L}^{2}\left|\sum_{\beta}\left[E\left(l_{b},l_{a},L\right)R^{\beta\beta}\left(ba;j_{L}\right)-iO\left(l_{b},l_{a},L\right)\beta R^{\beta-\beta}\left(ba;j_{L}\right)\right]\right|^{2},
\end{equation}

\begin{equation}\label{eq:r_sv_oneparticle}
\begin{aligned}
R_{L}^{\mathrm{sv}}\left(ba;k\right)=\sum_{l} & i^{L-l}\Pi_{lL}C_{L010}^{l0}\sum_{\beta}\left[E\left(l_{b},l_{a},L\right)R^{\beta\beta}\left(ba;j_{L}\right)+iO\left(l_{b},l_{a},L\right)\beta R^{\beta-\beta}\left(ba;j_{L}\right)\right]\\
\times & \left[E\left(l_{b},l_{a},l\right)\sum_{\beta}\beta S_{l}^{\beta}\left(\varkappa_{b},\varkappa_{a}|L\right)R^{\beta\beta}\left(ba;j_{l}\right)-iO\left(l_{b},l_{a},l\right)\sum_{\beta}U_{l}^{\beta}\left(\varkappa_{b},\varkappa_{a}|L\right)R^{\beta-\beta}\left(ba;j_{l}\right)\right],
\end{aligned}
\end{equation}

\begin{equation}\label{eq:r_vv1_oneparticle}
R_{L}^{\mathrm{vv},1}\left(ba;k\right)=\sum_{l}\Pi_{l}^{2}\left|\left[-E\left(l_{b},l_{a},l\right)\sum_{\beta}\beta S_{l}^{\beta}\left(\varkappa_{b},\varkappa_{a}|L\right)R^{\beta\beta}\left(ba;j_{l}\right)+iO\left(l_{b},l_{a},l\right)\sum_{\beta}U_{l}^{\beta}\left(\varkappa_{b},\varkappa_{a}|L\right)R^{\beta-\beta}\left(ba;j_{l}\right)\right]\right|^{2},
\end{equation}

\begin{equation}\label{eq:r_vv2_oneparticle}
\begin{aligned}
R_{L}^{\mathrm{vv},2}\left(ba;k\right)= & \sum_{l_{1}l_{2}}i^{-l_{1}+l_{2}}\Pi_{l_{1}l_{2}}C_{L010}^{l_{2}0}C_{L010}^{l_{1}0}\\
\times & \left[E\left(l_{b},l_{a},l_{1}\right)\sum_{\beta}\beta S_{l_{1}}^{\beta}\left(\varkappa_{b},\varkappa_{a}|L\right)R^{\beta\beta}\left(ba;j_{l_{1}}\right)-iO\left(l_{b},l_{a},l_{1}\right)\sum_{\beta}U_{l_{1}}^{\beta}\left(\varkappa_{b},\varkappa_{a}|L\right)R^{\beta-\beta}\left(ba;j_{l_{1}}\right)\right]\\
\times & \left[E\left(l_{b},l_{a},l_{2}\right)\sum_{\beta}\beta S_{l_{2}}^{\beta}\left(\varkappa_{b},\varkappa_{a}|L\right)R^{\beta\beta}\left(ba;j_{l_{2}}\right)+iO\left(l_{b},l_{a},l_{2}\right)\sum_{\beta}U_{l_{2}}^{\beta}\left(\varkappa_{b},\varkappa_{a}|L\right)R^{\beta-\beta}\left(ba;j_{l_{2}}\right)\right].
\end{aligned}
\end{equation}

Finally, the triple sum over the total angular-momentum projections of the particles $a$ and $b$ and the sum over intermediate angular momentum projections $M$ is evaluated using the sum rules for the relativistic Gaunt coefficients, Eq.~\eqref{eq:rgaunt_sumrule}.
The result is
\begin{equation}
\begin{aligned}\sum_{M\mu_{b}\mu_{a}}\int\,d\vec{k}\,L_{\alpha\beta}\left(ba;\vec{k}\right)\left(-g^{\alpha\beta}k^{2}+k^{\alpha}k^{\beta}\right) = 4\pi\Pi_{j_{b}}^{2}\sum_{L} & \left(C_{j_{b}\frac{1}{2}L0}^{j_{a}\frac{1}{2}}\right)^{2}\int_0^{k^0} dk\,k^{2}\Big\{ k^{2}R_{L}^{\mathrm{ss}}\left(ba;k\right)-2k^{0}k\mathrm{Re}R_{L}^{\mathrm{sv}}\left(ba;k\right)+\\
 & +\left[\left(k^{0}\right)^{2}-k^{2}\right]R_{L}^{\mathrm{vv},1}\left(ba;k\right)+k^{2}R_{L}^{\mathrm{vv},2}\left(ba;k\right)\Big\}.
\end{aligned}
\end{equation}

\onecolumngrid
\section{A relation between the two-particle and effective one-particle radial integrals}\label{app:g}
We start by equating the right-hand sides of Eqs.~\eqref{eq:e_spectrum_2body} and~\eqref{eq:e_spectrum_1body} and restrict ourselves with a comparison of the scalar contribution only.
For our purposes, the explicit form of the central-field radial wave functions for a massless particle, $E>0$, and antiparticle, $E<0$, are needed
\begin{equation}
F_{E\varkappa}^{\beta}\left(r\right)=\frac{1}{c}\sqrt{\frac{E^{2}}{\pi}} \times
\begin{cases}
rj_{l}\left(Er\right), & \beta=+1,\\
S_{E}S_{\varkappa_{d}}rj_{\overline{l}}\left(Er\right), & \beta=-1,
\end{cases}
\end{equation}
where $S_{x}= \mathrm{sgn}(x)$ and $\overline{l}=\left|-\varkappa+\frac{1}{2}\right|-\frac{1}{2}$.
It should be noted that the following relation between the radial components of the wave function for a massless particle holds
\begin{equation}\label{eq:radwf_relation_massless}
F_{E-\varkappa}^{\beta}\left(r\right)=S_{E}S_{\varkappa}\beta F_{E\varkappa}^{-\beta}\left(r\right).
\end{equation}
Without loss of generality, let us fix the states $a$, $c$, the $l$-th term of the multipole sum and consider the case $E\left(l,l_a,l_c\right)=1$.
We spin off Eq.~\eqref{eq:e_spectrum_2body} employing the definition of the two-particle radial integrals, Eqs.~\eqref{eq:two_particle_radints}, changing the order of the radial and energy integration, and pulling aside everything what depends on the state $a$ and $c$.
After these manipulations, Eq.~\eqref{eq:e_spectrum_2body} becomes proportional to
\begin{equation}\label{eq:g3}
\begin{aligned}\frac{dW^{2\mathrm{p}}\left(E_{e},n_{\mu}\varkappa_{\mu}\right)}{dE_{e}}\propto\frac{1}{16\pi}\int_{0}^{\Delta}dE_{d}\,\sum_{\varkappa_{d}\varkappa_{b}}\left(\Pi_{j_{d}}C_{l0j_{d}\frac{1}{2}}^{j_{b}\frac{1}{2}}\right)^{2} & \Bigg\{ \bigg[ E\left(l,l_{b},l_{d}\right)\sum_{\beta}\left(F_{E_{d}\varkappa_{d}}^{\beta}\left(r_{1}\right)\right)^{*}F_{-\left(\Delta-E_{d}\right)\varkappa_{b}}^{\beta}\left(r_{1}\right) \\
 & +iO\left(l,l_{b},l_{d}\right)\sum_{\beta}\beta\left(F_{E_{d}\varkappa_{d}}^{\beta}\left(r_{1}\right)\right)^{*}F_{-\left(\Delta-E_{d}\right)\varkappa_{b}}^{-\beta}\left(r_{1}\right)\bigg]\\
 & \times \bigg[E\left(l,l_{b},l_{d}\right)\sum_{\beta}\left(F_{E_{d}\varkappa_{d}}^{\beta}\left(r_{2}\right)\right)^{*}F_{-\left(\Delta-E_{d}\right)\varkappa_{b}}^{\beta}\left(r_{2}\right)\\
 & +iO\left(l,l_{b},l_{d}\right)\sum_{\beta}\beta\left(F_{E_{d}\varkappa_{d}}^{\beta}\left(r_{2}\right)\right)^{*}F_{-\left(\Delta-E_{d}\right)\varkappa_{b}}^{-\beta}\left(r_{2}\right)\bigg]^{*}\Bigg\}.
\end{aligned}
\end{equation}
With the aid of Eq.~\eqref{eq:radwf_relation_massless}, we note that Eq.~\eqref{eq:g3} is invariant under the change $\varkappa\to-\varkappa$.
Therefore, for each $|\varkappa|$ one can sum over the signs of $\varkappa$; this results leads to an additional factor of $4$.
On the other hand, from Eq.~\eqref{eq:e_spectrum_1body} we obtain
\begin{equation}
\frac{dW^{1\mathrm{p}}\left(E_{e},n_{\mu}\varkappa_{\mu}\right)}{dE_{e}}\propto\frac{1}{12\pi^{3}}\int_{0}^{\Delta}dk\,k^{4} j_{l}\left(kr_{1}\right)j_{l}^{*}\left(kr_{2}\right).
\end{equation}
\par
Let us introduce
\begin{equation}
j_{pq}\left(z,y;x\right)=j_{p}\left(yx\right)j_{q}\left(y\left(z-x\right)\right),
\end{equation}
Then, for $\alpha,\beta,\gamma>0$ and $j_{s}=s-\frac{1}{2}$ we obtain
\begin{equation}
\begin{aligned}\int_{0}^{\alpha}dx\,x^{4}j_{l}\left(\beta x\right)j_{l}\left(\gamma x\right)= & 3\sum_{pq}\left(\Pi_{j_{q}}C_{j_{q}\frac{1}{2}l0}^{j_{p}\frac{1}{2}}\right)^{2}\int_{0}^{\alpha}dx\,x^{2}\left(\alpha-x\right)^{2}\\
\times & \bigg\{ E\left(l,p,q\right)\left[j_{pq}\left(\alpha,\beta;x\right)-j_{p-1q-1}\left(\alpha,\beta;x\right)\right]\left[j_{pq}\left(\alpha,\gamma;x\right)-j_{p-1q-1}\left(\alpha,\gamma;x\right)\right]\\
+ & O\left(l,p,q\right)\left[j_{pq-1}\left(\alpha,\beta;x\right)+j_{p-1q}\left(\alpha,\beta;x\right)\right]\left[j_{pq-1}\left(\alpha,\gamma;x\right)+j_{p-1q}\left(\alpha,\gamma;x\right)\right]\bigg\}.
\end{aligned}
\end{equation}
The validity of this equation is confirmed numerically.
It is possible to further simplify this relation by using properties of the Clebsch-Gordan coefficients
\begin{equation}\label{eq:g7}
\begin{aligned} & \int_{0}^{\alpha}dx\,x^{4}j_{l}\left(\beta x\right)j_{l}\left(\gamma x\right)=\frac{3}{\left(2l+1\right)}\sum_{pq}\int_{0}^{\alpha}dx\,x^{2}\left(\alpha-x\right)^{2}\\
\times & \Bigg\{\left(p+q-l\right)\left(p+q+l+1\right)\left(C_{q0\,p0}^{l0}\right)^{2}\left[j_{pq}\left(\alpha,\beta;x\right)-j_{p-1q-1}\left(\alpha,\beta;x\right)\right]\left[j_{pq}\left(\alpha,\gamma;x\right)-j_{p-1q-1}\left(\alpha,\gamma;x\right)\right]\\
 & -\left(p-q-l\right)\left(p-q+l+1\right)\left(C_{q-10\,p0}^{l0}\right)^{2}\left[j_{pq-1}\left(\alpha,\beta;x\right)+j_{p-1q}\left(\alpha,\beta;x\right)\right]\left[j_{pq-1}\left(\alpha,\gamma;x\right)+j_{p-1q}\left(\alpha,\gamma;x\right)\right]\Bigg\}.
\end{aligned}
\end{equation}
Using the expansion of Eq.~\eqref{eq:g7} as $\gamma\to0$ and the asymptotic behavior of the spherical Bessel function,
\begin{equation}
j_{n}\left(z\right)\stackrel{z\to0}{\sim}\frac{z^{n}}{\left(2n+1\right)!!},
\end{equation}
yields
\begin{equation}
\begin{aligned} & \int_{0}^{\alpha}dx\,x^{l+4}j_{l}\left(\beta x\right)=-3\left(2l-1\right)!!\sum_{pq}\int_{0}^{\alpha}dx\,\frac{x^{p+1}\left(\alpha-x\right)^{q+1}}{\left(2p-1\right)!!\left(2q-1\right)!!}\\
\times & \Bigg\{\left(p+q-l\right)\left(p+q+l+1\right)\left(C_{q0\,p0}^{l0}\right)^{2}\left[j_{pq}\left(\alpha,\beta;x\right)-j_{p-1q-1}\left(\alpha,\beta;x\right)\right]\delta_{p+q,l+2}\\
 & +\left(p-q-l\right)\left(p-q+l+1\right)\left(C_{q-10\,p0}^{l0}\right)^{2}\left[j_{pq-1}\left(\alpha,\beta;x\right)+j_{p-1q}\left(\alpha,\beta;x\right)\right]\left(\frac{x}{2p+1}+\frac{\alpha-x}{2q+1}\right)\delta_{p+q,l+1}\Bigg\}.
\end{aligned}
\end{equation}
An additional expansion in $\beta$ results in relations for sums of the Clebsch-Gordan coefficients:
\begin{equation}
\begin{aligned}\frac{1}{2l+1}\frac{\left(2l+4\right)!}{\left[\left(2l-1\right)!!\right]^{2}} & =6\sum_{pq}\frac{\left(2q\right)!\left(2p\right)!}{\left[\left(2q-1\right)!!\left(2p-1\right)!!\right]^{2}}\\
 & \times\left[\left(2l+3\right)\left(C_{q0\,p0}^{l0}\right)^{2}\delta_{p+q,l+2}+2p\left(2q-1\right)\left(2+\frac{1}{2p+1}\right)\left(C_{q-10\,p0}^{l0}\right)^{2}\delta_{p+q,l+1}\right].
\end{aligned}
\end{equation}

\end{document}